\documentclass[a4paper,fleqn,usenatbib]{mnras}

\usepackage[T1]{fontenc}
\usepackage{ae,aecompl}
\usepackage{multirow}

\usepackage{graphicx}	
\usepackage{amsmath}	
\usepackage{amssymb}	

\title[Photometric redshifts for radio surveys - II]{Photometric redshifts for the next generation of deep radio continuum surveys - II. Gaussian processes and hybrid estimates}

\author[K. J. Duncan et al.]{Kenneth J Duncan$^{1}$\thanks{E-mail: duncan@strw.leidenuniv.nl},
Matt J. Jarvis$^{2,3}$,
Michael J. I. Brown$^{4,5}$, \newauthor
Huub J. A. R\"{o}ttgering$^{1}$
\\
$^{1}$Leiden Observatory, Leiden University, NL-2300 RA Leiden, Netherlands \\
$^{2}$Astrophysics, University of Oxford, Denys Wilkinson Building, Keble Road, Oxford, OX1 3RH \\
$^{3}$Physics and Astronomy Department, University of the Western Cape, Bellville 7535, South Africa\\
$^{4}$School of Physics and Astronomy, Monash University, Clayton, Victoria 3800, Australia \\
$^{5}$Monash Centre for Astrophysics, Monash University, Clayton, Victoria, 3800, Australia \\
}

\pubyear{2018}

\begin{document}
\label{firstpage}
\pagerange{\pageref{firstpage}--\pageref{lastpage}}
\maketitle

\defcitealias{Duncan:2017wu}{Paper I}

\begin{abstract}
Building on the first paper in this series \citep{Duncan:2017wu}, we present a study investigating the performance of Gaussian process photometric redshift (photo-$z$) estimates for galaxies and active galactic nuclei detected in deep radio continuum surveys.
A Gaussian process redshift code is used to produce photo-$z$ estimates targeting specific subsets of both the AGN population - infrared, X-ray and optically selected AGN - and the general galaxy population.
The new estimates for the AGN population are found to perform significantly better at $z > 1$ than the template-based photo-$z$ estimates presented in our previous study.
Our new photo-$z$ estimates are then combined with template estimates through hierarchical Bayesian combination to produce a hybrid consensus estimate that outperforms both of the individual methods across all source types.
Photo-$z$ estimates for radio sources that are X-ray sources or optical/IR AGN are significantly improved in comparison to previous template-only estimates - with outlier fractions and robust scatter reduced by up to a factor of $\sim4$.
The ability of our method to combine the strengths of the two input photo-$z$ techniques and the large improvements we observe illustrate its potential for enabling future exploitation of deep radio continuum surveys for both the study of galaxy and black hole co-evolution and for cosmological studies.

\end{abstract}




\section{Introduction}
Photometric redshifts (photo-$z$s hereafter) have become a fundamental tool for both the study of galaxy evolution and for modern cosmology experiments.
The main driving force behind recent developments in photometric redshift estimation methodology has been the stringent requirements set by the coming generation of weak-lensing cosmology experiments \citep[e.g. {\em EUCLID};][]{2011arXiv1110.3193L}.
However, the need for accurate and unbiased redshift estimates for large samples of galaxies ($\approx 10^{6}-10^{9}$) represents a near universal requirement for all future extra-galactic surveys.

Through either template based \citep[e.g.][]{Arnouts:1999gh,Bolzonella:2000uw,Benitez:2000jr,Brammer:2008gn} or empirical/`machine learning' \citep[e.g.][]{Collister:2004fx,Geach:2011fn,CarrascoKind:2013kd,CarrascoKind:2014gb} estimation techniques, it is now possible to produce the precise and reliable photometric redshifts required for optically selected galaxy samples \citep{Bordoloi:2010df, Sanchez:2014gf, CarrascoKind:2014jg, 2017arXiv170801531D}.
However, typically such methods are applied to, or optimised for, the galaxy emission due to stellar populations, with galaxies dominated by emission from active galactic nuclei (AGN) either removed from the analysis (where possible) or not explicitly accounted for.
This therefore presents a problem in surveys where a larger fraction of the population is composed of AGN, for example in radio-continuum selected surveys \citep[and for the $\sim3$ million X-ray selected AGN and QSOs observed by the \textit{eRosita} mission, see][]{2012arXiv1209.3114M}.
The population of radio detected sources is extremely diverse - with radio emission tracing both black hole accretion in AGN and star formation activity.

Probing to unprecedented depths, deep radio continuum surveys from MeerKAT \citep{Booth:2009wx}, the Australian SKA Pathfinder \citep[ASKAP;][]{2007PASA...24..174J} and the Low Frequency Array \citep[LOFAR;][]{vanHaarlem:2013gi} will increase the detected population of radio sources by more than an order of magnitude and probe deep into the earliest epochs of galaxy formation and evolution \citep{2010iska.meetE..50R, Norris:2013jo, 2017arXiv170901901J}.
 Accurate and unbiased photometric redshift estimates for the full radio source population will be essential for studying the faint radio population and achieving the scientific goals of these deep radio continuum surveys - both for galaxy evolution and cosmological studies.

In \citet[][hereafter Paper I]{Duncan:2017wu}, we investigated the performance of template-based photometric redshift estimates for the radio-continuum detected population over a wide range of optical and radio properties. 
Specifically, three photometric redshift template sets, representative of those available in the literature, were applied to two optical/IR datasets and their performance investigated as a function of redshift, radio flux/luminosity and infrared/X-ray properties.

Furthermore, by combining all three photo-$z$ estimates through hierarchical Bayesian combination \citep{Dahlen:2013eu,CarrascoKind:2014jg} we were able to produce a new consensus estimate that outperforms any of the individual estimates that went into it.
Although the consensus redshift estimates were found to offer some improvement, the overall quality of template photo-$z$ estimates for radio sources that are X-ray sources or optical/IR AGN was still relatively poor.
The measured outlier fractions and scatter relative to the spectroscopic training sample remained unacceptable for some science goals, including multi-messenger cosmological studies \citep{2012MNRAS.427.2079C,2014MNRAS.442.2511F, 2015aska.confE..18J}, radio weak lensing experiments \citep{Brown:2015vu} and galaxy/AGN evolution studies that rely on optical quasar samples \citep{2017MNRAS.469.1883M}.
An alternative methodology is therefore needed to either replace the template-based photo-$z$ estimates for these difficult populations or help to improve the consensus estimate.

Empirical (or machine learning) photo-$z$ estimates have already been shown to offer a potential solution for improving photo-$z$s for the AGN population \citep[e.g.][]{Richards:2001ct, Brodwin:2006dp, Bovy:2012gj}.
In this paper we investigate how such machine learning photo-$z$ techniques perform when applied to the same samples and data where template-based methods were found to struggle the most in \citetalias{Duncan:2017wu}. 
Specifically, we explore the use of Gaussian processes (GP) using the framework presented by \citet[][GPz]{2016MNRAS.455.2387A, 2016MNRAS.462..726A}.
GPz offers several key advantages that make it an ideal choice for tackling the problems posed by large samples of radio selected galaxies. 
Firstly, it has been shown to outperform other empirical photo-$z$ tools in the literature when applied to sparse datasets. 
Secondly, it incorporates cost-sensitive learning, i.e. the ability to give more or less weight to certain sources during the optimisation procedure. 
These additional weights potentially allow for biases in the available training sample to be accounted for. 
Finally, by modelling the non-uniform noise intrinsic in photometric datasets it offers estimations of the variance on the predicted photo-$z$s - meaning that its outputs can also be easily incorporated into the hierarchical Bayesian combination framework presented in \citetalias{Duncan:2017wu}.

This paper is organized as follows: 
Section~\ref{sec:data} presents the data used in this study along with details of the multi-wavelength classifications employed throughout the work.
Section~\ref{sec:deepfields} then outlines the application of the \textsc{GPz} framework to photometric data from deep survey fields such as those explored in \citetalias{Duncan:2017wu} and the improvements that can be made in photometric redshift qualilty for the most difficult radio source populations.
In Section~\ref{sec:combined}, we present the results of incorporating the new GP photo-$z$s within the Bayesian combination framework presented in \citetalias{Duncan:2017wu}.
Finally, Section~\ref{sec:summary} presents a brief summary of the results in this paper and the key conclusions we draw. 

Throughout this paper, all magnitudes are quoted in the AB system \citep{1983ApJ...266..713O} unless otherwise stated. We also assume a $\Lambda$-CDM cosmology with $H_{0} = 70$ kms$^{-1}$Mpc$^{-1}$, $\Omega_{m}=0.3$ and $\Omega_{\Lambda}=0.7$.

\section{Data}\label{sec:data}
In \citetalias{Duncan:2017wu} we made use of two samples of galaxies drawn from both a wide area optical survey \citep[NDWFS Bo\"{o}tes;][]{Jannuzi:1999wu} and a smaller but deeper optical survey field \citep[COSMOS;][]{Laigle:2016ku}.
Although we apply the method outlined in the following section to both samples, in this paper we will concentrate mainly on the `Wide' field sample in our subsequent analysis.
The reasons for this are two-fold:
Firstly, the targeted selection criteria of the AGN and Galaxy Evolution Survey \citep[AGES;][]{Kochanek:jy} spectroscopic survey in the field results in a larger sample of AGN sources (see Fig.1 of \citetalias{Duncan:2017wu}) for training and testing the GP redshift estimates.
Although the overall spectroscopic training sample available in COSMOS is larger than that of Bo\"{o}tes, the number of training sources available for some subsets of the AGN population (IR and optically selected) is lower by up to a factor of four.

Secondly, the optical filter coverage and depth of the available photometry in the field is more representative of the large optical survey fields that are being observed with deep radio continuum surveys such as LOFAR.
Furthermore, the poorer quality of AGN template estimates in the `Wide' data is such that these datasets are where the desired improvements are greatest.
Nevertheless, we apply the method to both samples and the results for the `Deep' field are summarised and discussed with respect to the `Wide' field in Section~\ref{sec:deep}.

We refer the reader to \citetalias{Duncan:2017wu} and references therein for full details on the Bo\"{o}tes photometric catalog itself, along with details on the spectroscopic redshift information available in the field.
As in \citetalias{Duncan:2017wu}, the radio continuum observations from this field are taken from the LOFAR observations presented in \citet{Williams:2016bf}.
Details of the cross-matching procedure between the radio data and the optical catalog used in this work can be found in \citet{2018MNRAS.475.3429W}.

Given its importance in the subsequent analysis it is worth summarising the multi-wavelength AGN classifications applied to the data.
We classify all sources in the Bo\"{o}tes spectroscopic comparison samples using the following additional criteria:
\begin{itemize}
\item \emph{Infrared AGN} are identified using the updated IR colour criteria presented in \citet{Donley:2012ji}.
\item \emph{X-ray AGN} in the Bo\"{o}tes field were identified by cross-matching the positions of sources in our catalog with the X-B\"{o}otes \emph{Chandra} survey of NDWFS \citep{Kenter:2005gj}.
We calculate the x-ray-to-optical flux ratio, $X/O = \log_{10}(f_{X}/f_{\textup{opt}})$, based on the $I$ band magnitude following \citet{Brand:2006iv} and for a source to be selected as an X-ray AGN, we require that an x-ray source has $X/O > -1$ or an x-ray hardness ratio $> 0.8$ \citep{2004AJ....128.2048B}.
\item \emph{Optical AGN} were also identified through cross-matching the optical catalog with the Million Quasar Catalog compilation of optical AGN, primarily based on SDSS \citep{2015ApJS..219...12A} and other literature catalogs \citep{2015PASA...32...10F}. 
\end{itemize}

\begin{figure}
\centering
	\includegraphics[width=0.5\columnwidth]{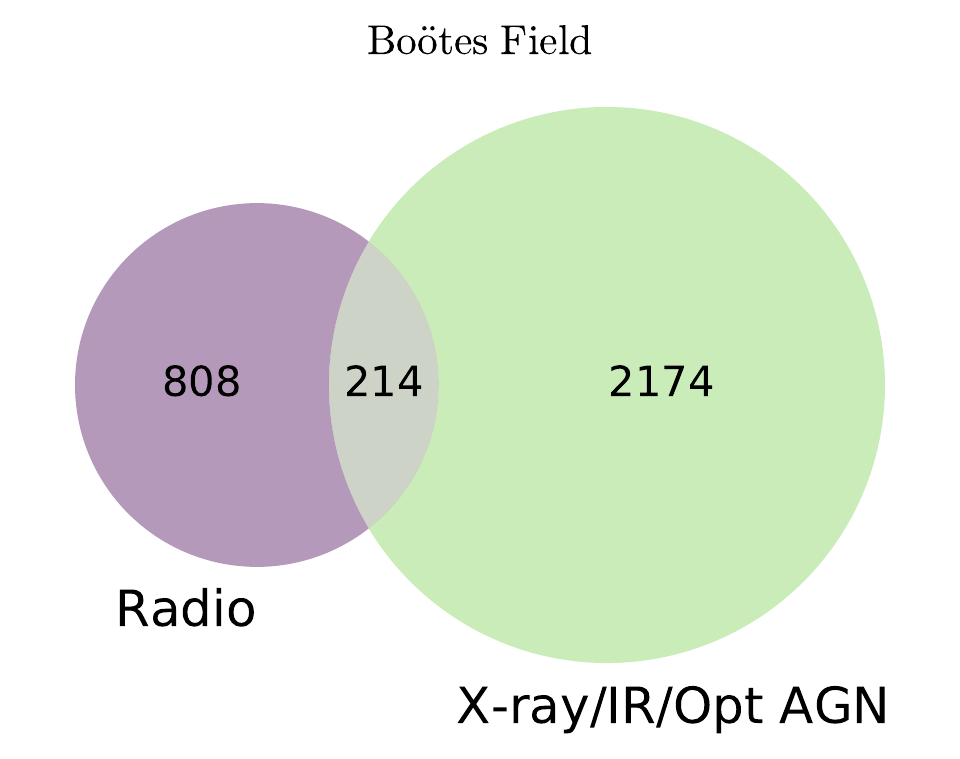} \includegraphics[width=0.4\columnwidth]{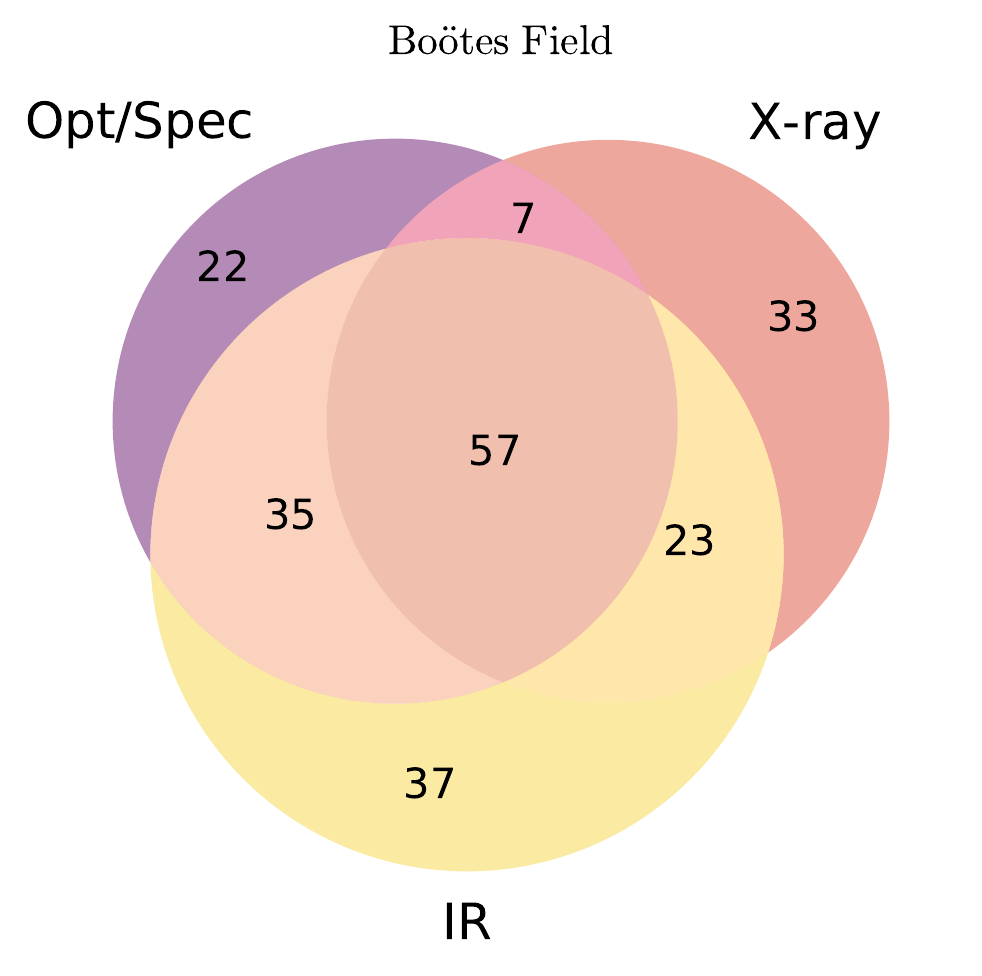}
  \caption{Multi-wavelength classifications of the sources in the full spectroscopic redshift sample for the Bo\"{o}tes dataset used in this study. The `Radio' and  `X-ray/IR/Opt AGN' subsets correspond respectively to radio detected sources and identified X-ray sources and optical/spectroscopic/infra-red selected AGN (see Section~\ref{sec:data}). As illustrated in previous studies, the X-ray, IR AGN and radio source population are largely distinct populations with only partial overlap.}
  \label{fig:venn}
\end{figure}

Note however, these classifications are not expected to be distinct physical classifications but rather selection methods through which a wide variety of the most luminous AGN can be identified.
Depending on data available in a given field, further sub-classifications or alternative criteria might be warranted.
As shown in Fig.~\ref{fig:venn}, there is significant overlap between different selection criteria with the majority of radio sources selected as AGN belonging to at least two of the subsets.
Despite these overlaps, there is also potentially a very wide variety of intrinsic spectral energy distributions within the full AGN sample, both between these subsets of AGN and within the subsets themselves.

As in \citetalias{Duncan:2017wu}, spectroscopic redshifts for sources in Bo\"{o}tes are taken from a compilation of observations within the field comprising primarily of the results of the AGN and Galaxy Evolution Survey \citep[AGES;][]{Kochanek:jy} spectroscopic survey, with additional redshifts provided by a large number of smaller surveys in the field including \citet{2012ApJ...758L..31L,2013ApJ...771...25L,2014ApJ...796..126L}, \citet{2012ApJ...753..164S}, \citet{2012ApJ...756..115Z,2013ApJ...779..137Z} and \citet{2016ApJ...823...11D}.

In total, the combined sample consists of 22830 redshifts over the range $0 < z < 6.12$, with 88\% of these at $z < 1$.
Due to the nature of the AGES target selection criteria, identified AGN sources have a higher degree of spectroscopic completeness than the general galaxy population ($\approx 11\%$ of AGN have spectroscopic redshifts available compared to $\approx 1\%$ of the rest of the galaxy population).
Nevertheless, as is the case in most spectroscopic training samples the available sources do not necessarily sample the full photometric colour space. 
In the following section we present the weighting strategy employed to minimise the potential effects caused by the biased training sample. The limitations of the training sample and ways in which this can be mitigated in the future will also be revisited in Section~\ref{sec:future}.

\section{Gaussian Process photometric redshifts for AGN in deep fields}\label{sec:deepfields}
The use of Gaussian processes (GP) for regression \citep{2006gpml.book.....R} has become increasingly popular in recent years, primarily due to its advantages of being a Bayesian method that is both non-linear and non-parametric.
The Gaussian process photometric redshift code, GPz \citep{2016MNRAS.455.2387A} extends the standard GP method to add several features suited to photo-$z$ estimation.
Firstly, \citet{2016MNRAS.455.2387A} introduces sparse GPs that lower the computational requirements without significantly affecting accuracy of the models.
Secondly, \citet{2016MNRAS.462..726A} extends the method to account for non-uniform and variable noise (heteroscedastic) within the input data - modelling both the intrinsic noise within the photometric data and model uncertainties due to limited training data.
Finally, the code incorporates the option for cost-sensitive learning, allowing the weights of different parts of parameter space to be varied in order to optimise the analysis for a specific science goal.

Given a training set of input magnitudes and corresponding uncertainties, GPz models the distribution of functions that map those inputs onto the desired output (in this case, the spectroscopic redshift).
After optimisation, the model can then be used to predict the redshift and corresponding uncertainties (consisting of both noise and model components) for a new set of inputs. 
Detailed descriptions of the theoretical background and methodology of \textsc{GPz} are presented in \citet{2016MNRAS.455.2387A} and \citet{2016MNRAS.462..726A}.
In this section, we therefore outline only the details of how \textsc{GPz} was applied to our dataset.

\subsection{GPz Method}\label{sec:method}
Although the three different AGN selection criteria outlined in Section~\ref{sec:data} contain significant overlap in their populations, we choose to train and calibrate the GP estimates of each subset separately.

Due to both inhomogeneity in the coverage of different filters and the relatively shallow depth of some of these observations in the Bo\"{o}tes dataset, only a small fraction of sources are detected in all of the filters available in the field.
For example, only $\approx 9\%$ of the full Bo\"{o}tes photometric catalog has magnitude values available in the 13-bands extending from $u$-band to IRAC $8\mu\textup{m}$.
The number and combination of magnitudes input to \textsc{GPz} for each subset were therefore chosen to cover as broad a wavelength as possible whilst trying to ensure as many sources as possible were detected in the corresponding bands.
Starting with the detection band of the multi-wavelength catalog ($I$), additional filter choices were added and the fraction of sources with magnitudes available in those filters calculated until the fraction fell to $\sim 80\%$.
For cases where several different filter combinations offer a similar number of available sources, the combination that produces the best estimates in limited trials is chosen.
We note however that systematic searches for the best filter combinations have not been performed. 
We also note that an extension to GPz is being developed to account for missing data in a fully consistent way (Almosallam et al. in prep) such that these issues will be further minimised in future.

For the purposes of training each GPz classifier, each input sample was split at random into training, validation and test samples consisting of 80, 10 and 10\% of the full sample respectively.
Note, all statistics reported in Section~\ref{sec:method} are for only the test sample, which is not included in the training in any way.
The filter selections and the sizes of the corresponding Bo\"{o}tes training samples are as follows:
\begin{itemize}
	
\item \emph{Infrared AGN} -- For the subset of IR AGN, the input dataset includes the optical $R$ and $I$ magnitudes in addition to the four IRAC magnitudes used in the colour selection of the subset.
In the spectroscopic training set and full photometric IR AGN subsets, 98.9\% and 82.6\% of sources respectively have magnitudes in these bands.
Of the 1751 spectroscopic sources classified as IR AGN, the final training, validation and test samples therefore consist of 1385, 173 and 173 sources respectively.

\item \emph{X-ray AGN} -- The final filter choice for the X-ray AGN sources is $B_{w}$, $R$, $I$, $K_{s}$ and Spitzer/IRAC $3.6$ and $4.5\mu\textup{m}$.
Detection fractions in the spectroscopic and full photometric samples are almost identical to the IR AGN subset, with fractions of 98.8\% and 82.7\% respectively.
There are 1133 spectroscopic sources classified as X-ray AGN, resulting in training, validation and test samples of 895, 112 and 112 respectively.

\item \emph{Optical AGN} -- Although optically bright by definition, the chosen filter selection for the optical AGN subset consists of $I$ in combination with the near and mid-infrared bands of $J$, $Ks$, Spitzer/IRAC $3.6$/$4.5\mu\textup{m}$ and Spitzer/MIPS $24\mu\textup{m}$.
In these filters, the available training and full sample fractions are 96.6\% and 84.2\% respectively.
For the 1382 optical AGN sources in the spectroscopic training sample, this results in 1067, 134 and 134 sources in the training, validation and test samples. 
\end{itemize}

In addition to the three \textsc{GPz} estimators targeted at subsets of the AGN population, we also produce an additional estimator trained on optical sources that do not satisfy any of the AGN selection criteria - corresponding to the significant majority of both the training sample and photometric catalog.
As illustrated in the bottom panel of Fig.~\ref{fig:weights} (dashed blue line), the magnitude distribution for the full `galaxy' sample extends to significantly fainter magnitudes than those in the AGN subsets.
To find the optimum combination of optical bands we systematically calculated the fraction of sources with measured magnitudes in every possible combination of five bands out of those available in the field.
The two sets of filters that would allow estimates for the largest fraction of catalog sources are $\{$$u$, $B_{w}$, $R$, $I$, $z$$\}$ and $\{$$B_{w}$, $R$, $I$, $z$, $3.6\mu\textup{m}$$\}$, with 38.3\% and 34.2\% of the full photometric catalog respectively (87.3\% and 92.8\% of the training samples).

In all four cases, \textsc{GPz} was trained using 25 basis functions and allowing variable covariances for each basis function \citep[i.e. the `GPVC' of][]{2016MNRAS.455.2387A}.
We choose these parameters based on the tests of \citet{2016MNRAS.455.2387A} who found minimal performance gain above 25 basis functions and significant improvements when using fully variable covariances compared to other assumptions.
Finally, we also follow the practices in outlined Section 6.2 of \citet{2016MNRAS.455.2387A} by pre-processing the input data to normalise the data and de-correlate the features  (also known as `sphering' or `whitening').

\subsection{Weighting scheme}\label{sec:weights}
\begin{figure}
\centering
	\includegraphics[width=1\columnwidth]{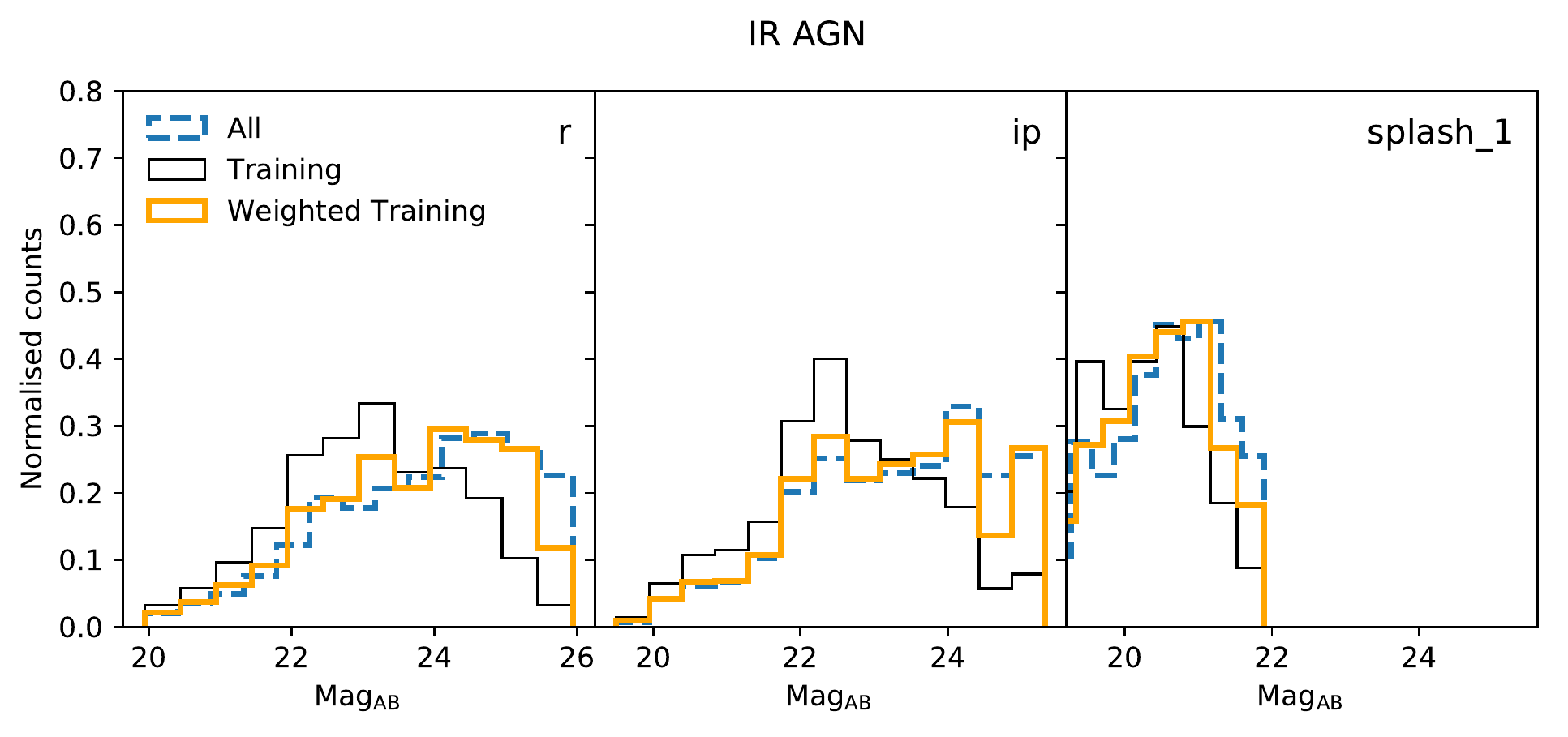}
	\includegraphics[width=1\columnwidth]{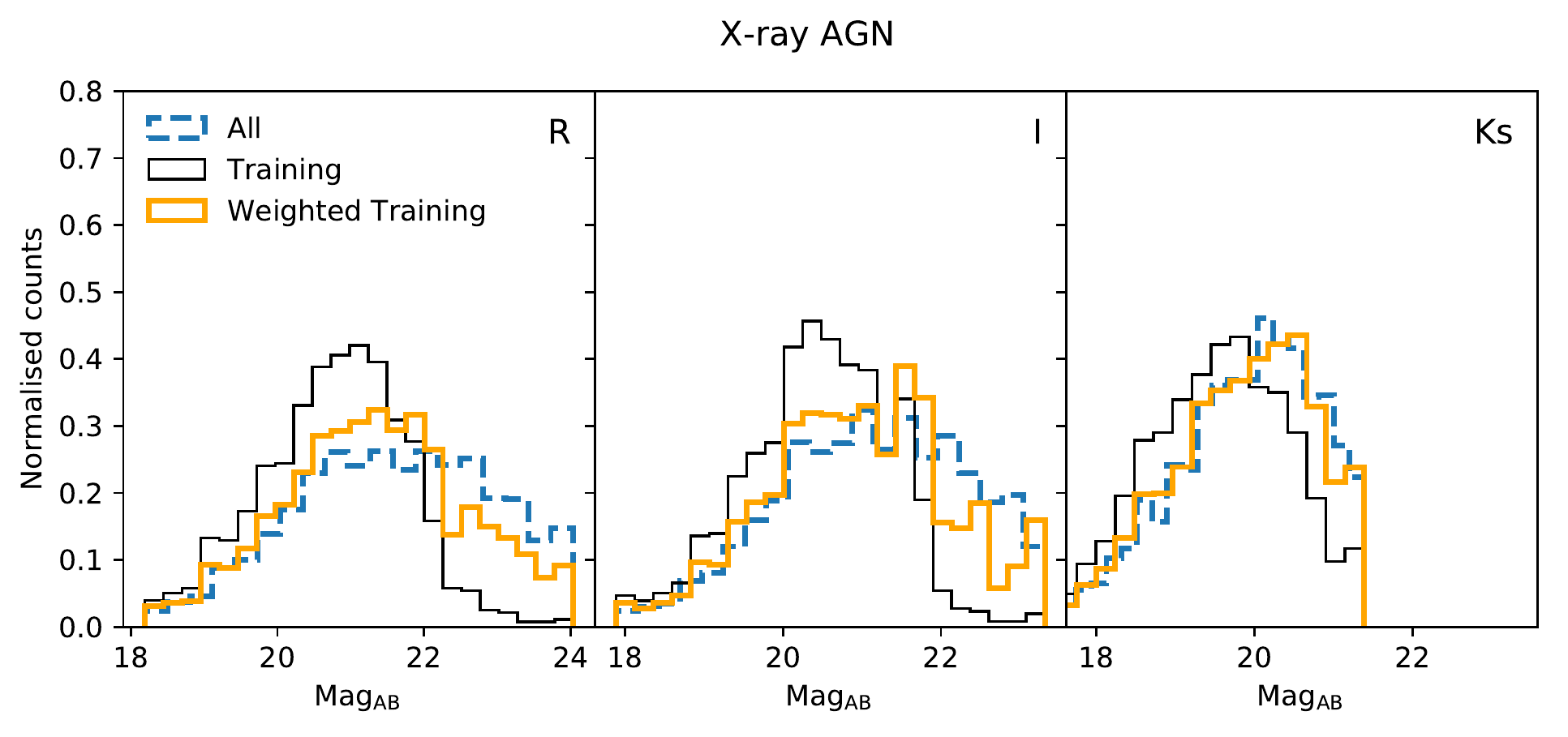} 
	\includegraphics[width=1\columnwidth]{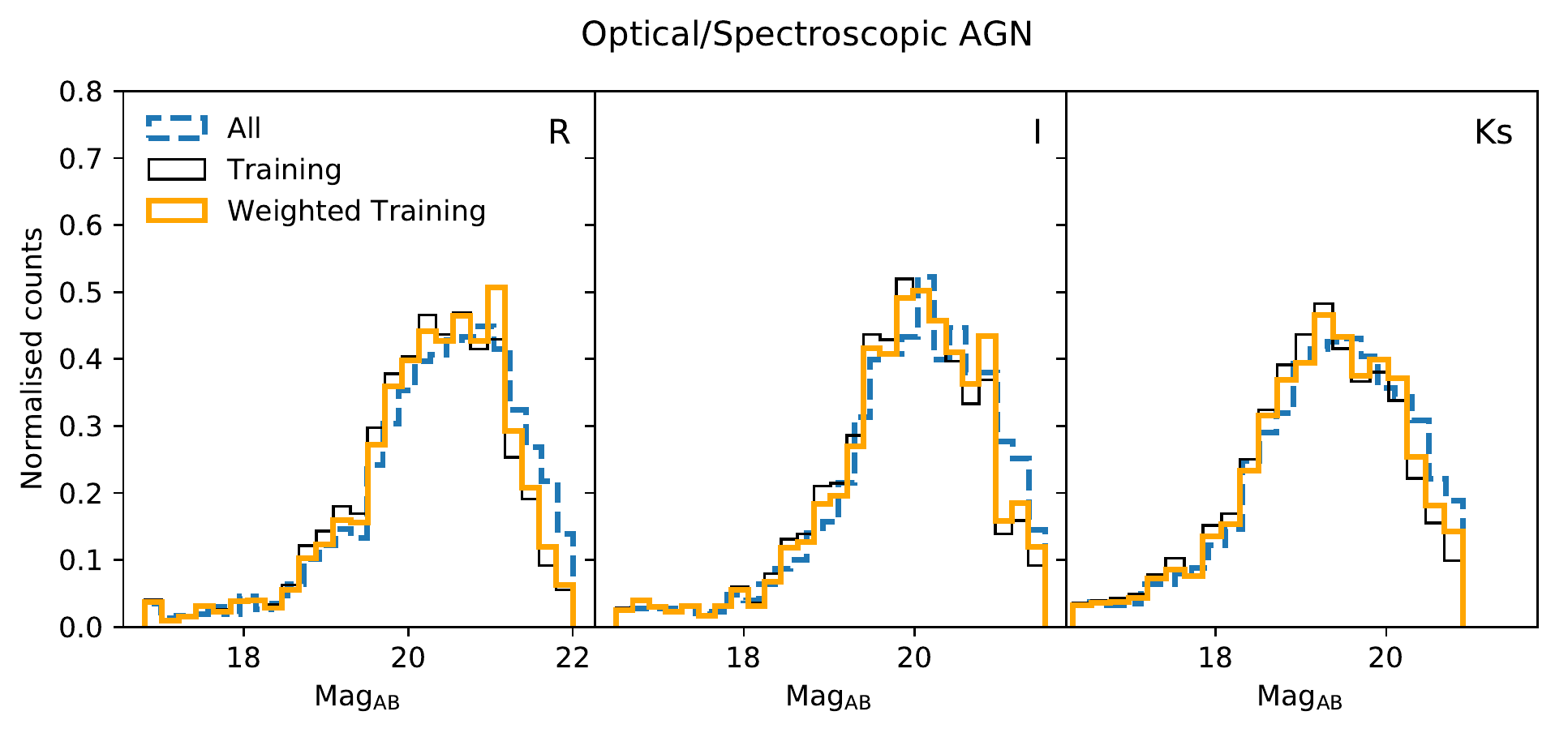} 
	\includegraphics[width=1\columnwidth]{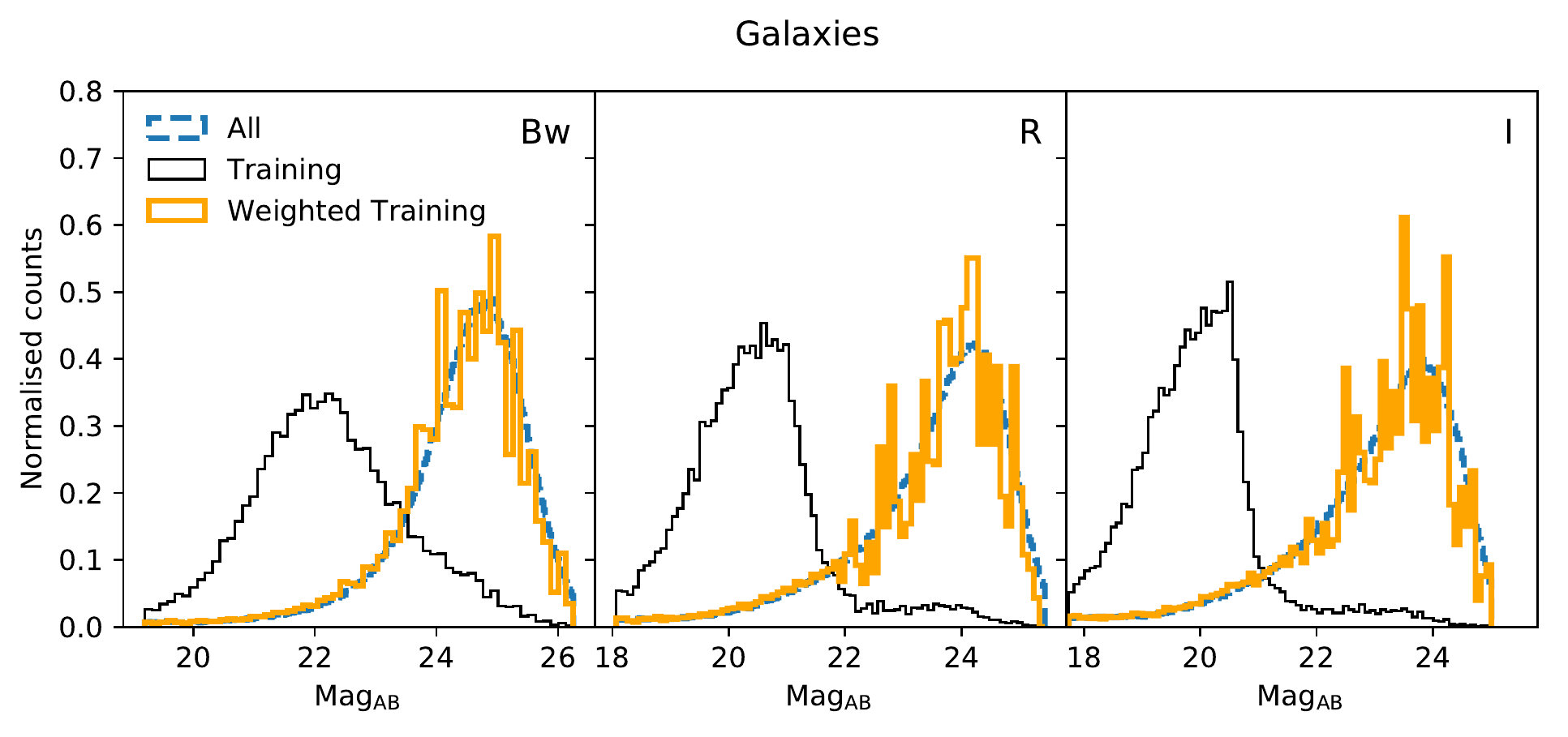} 

  \caption{Illustration of the colour-magnitude based weighting scheme applied to each of the Bo\"{o}tes field training subsets employed in this work. In each plot, the dashed blue line shows the magnitude distributions for the full photometric sample while the thin black and thick gold lines show the training sample before and after weighting. The optical/infrared filter corresponding to each magnitude distribution is labelled in the upper right corner of each plot - `ch1' and `ch2' correspond to the Spitzer/IRAC $3.6\mu\textup{m}$ and $4.5\mu\textup{m}$ filters respectively.}
  \label{fig:weights}
\end{figure}

One of the key advantages offered by \textsc{GPz} with respect to some other empirical methods in the literature is its option of using cost-sensitive learning; allowing for potential biases in the training sample to be taken into account or certain regions of parameter space to be prioritised if desired.
In this work we make use of two different weighting schemes. 
As a reference we first employ a flat weighting scheme \citep[i.e. the `Normal' weighting of][]{2016MNRAS.455.2387A}.
Secondly, we employ a weighting scheme that takes into account the colour and magnitude distribution of the training sample with respect to the full corresponding photometric sample.

Our colour based weighting scheme is based on the method presented in \citet{Lima:2008eu} and successfully employed elsewhere in the photo-$z$ literature \citep[e.g.][]{Sanchez:2014gf}.
Firstly, for all galaxies in the spectroscopic training set and the photometric sample we construct separate arrays consisting of the normalised distribution of $I$-band magnitudes and two photometric colours.
The colour and magnitude distributions are both normalised based on the 99th percentile range observed in the full photometric sample.
This renormalisation ensures that each observable is given equal importance in the subsequent weighting scheme and that the distribution is not severely affected by outliers.

Next, for each galaxy, $i$, in the spectroscopic training set, we compute the distance to the 9th nearest neighbour, $r_{i, 9}$, in the colour-magnitude space of the training set\footnote{The 9th nearest neighbour was chosen to provide marginally more localisation in the colour-magnitude space than the 16th nearest neighbour chosen in \citet{Lima:2008eu} while still minimising the effects of small-number statistics. However, as illustrated by the minimal effect on results for $4 < n < 64$ \citep{Lima:2008eu}, we do not expect this choice to have any significant effect on the results presented.}
We then find the corresponding number of objects, $N_{\textup{P}}(\textbf{m}_{i})$, in the full photometric sample that fall within a volume with radius equal to $r_{i, 9}$.
The weight for a given training galaxy, $W_{i}$, is then defined following Equation 24 of \citet{Lima:2008eu} such that
\begin{equation}
W_{i} = \frac{1}{N_{\textup{P}, \textup{tot}}} \frac{N_{\textup{P}}(\textbf{m}_{i})}{N_{\textup{T}}(\textbf{m}_{i})},
\end{equation}
where $N_{\textup{T}}(\textbf{m}_{i})$ is the number of objects in the training sample within the same volume (by definition 8 in this work) and $N_{\textup{P}, \textup{tot}}$ the total number of objects in the photometric training sample.
Finally, any training-set object with zero weight is removed from the sample and the weights renormalised such that $\sum_{i} W_{i} = 1$, to meet the convention required by GPz.

In Fig.~\ref{fig:weights} we illustrate the results of this weighting scheme for each of the training sample subsets used in our analysis.
For the three magnitudes used in the weighting scheme, Fig.~\ref{fig:weights} shows the magnitude distribution of the full photometric sample compared to that of the training sample before and after the weighting scheme has been applied.

The bias within the training sample is clearly strongest for both the IR AGN and normal galaxy populations, with the majority of training galaxies significantly brighter than those in the full photometric samples.
In both cases, the weighting scheme does a good job of reproducing the distribution of the full photometric sample.
However, as there are very few spectroscopic redshifts available at the very faintest optical magnitudes, the weighted training sample becomes somewhat noisy due to the small number of faint training objects being assigned high weights. 
Possible methods of minimising the effects of very small samples of faint training objects will be discussed further in Section~\ref{sec:future}.

\subsection{GPz photo-$z$ Results}
\begin{figure*}
\centering
	\includegraphics[width=1.5\columnwidth]{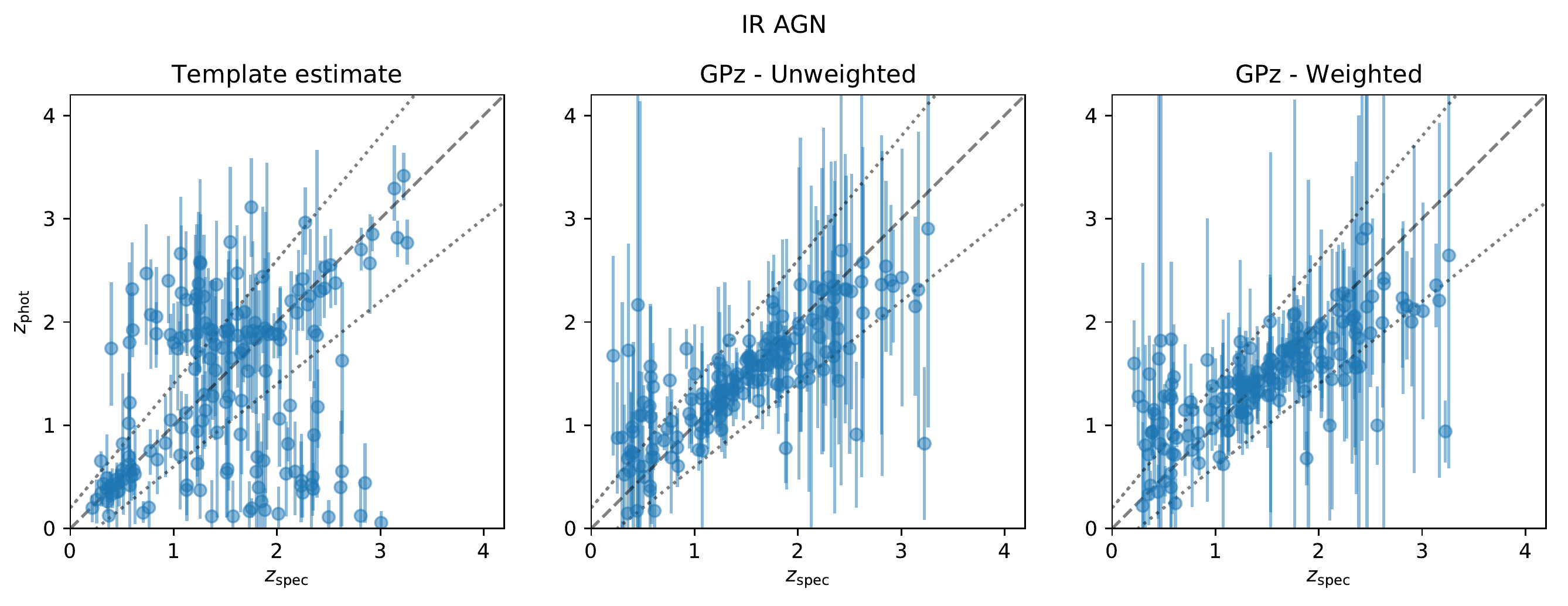} 
	\includegraphics[width=1.5\columnwidth]{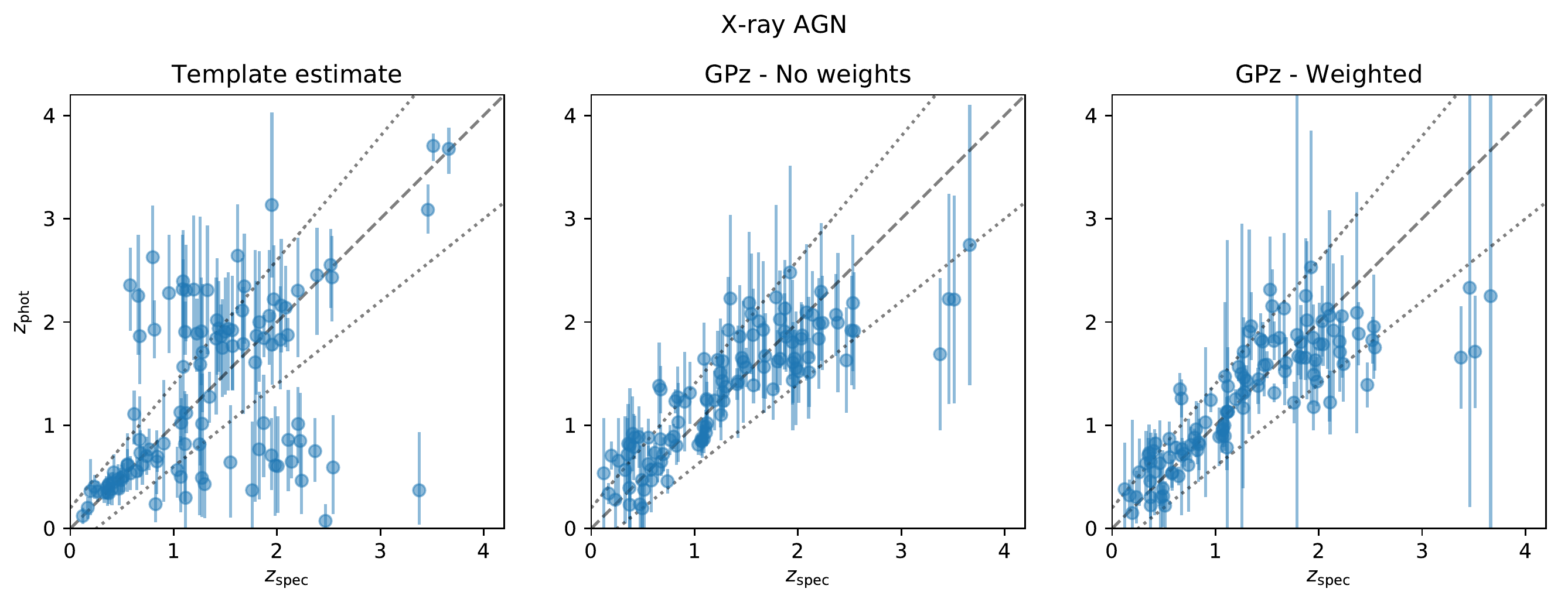} 
	\includegraphics[width=1.5\columnwidth]{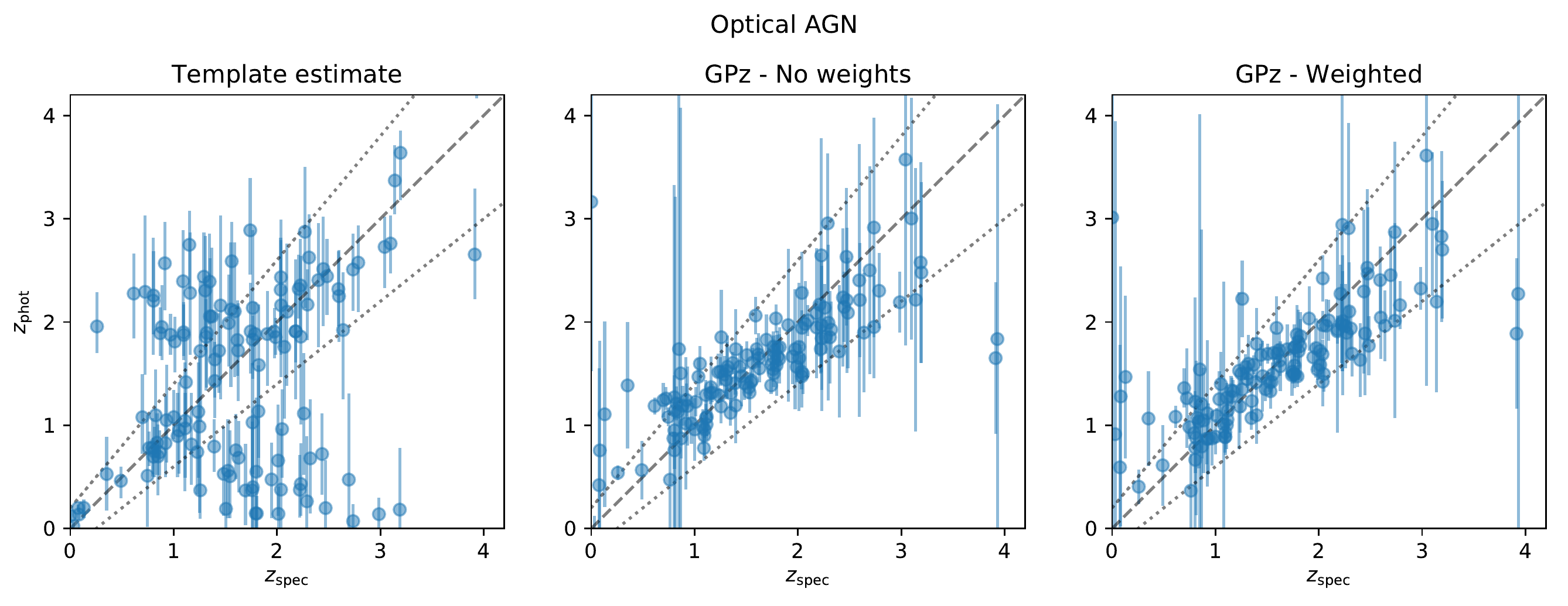} 

  \caption{Comparison of photometric redshift estimates versus the spectroscopic redshifts for each of the three Bo\"{o}tes AGN population subsets. The left column shows the consensus template-based photo-$z$ as calculated in Paper I. The centre and right-hand columns shows the results from the Gaussian process estimates when trained using the flat and colour-based weighting schemes respectively. The dashed grey line corresponds to the 1:1 relation while the dotted lines correspond to the outlier definition adopted in this work.}
  \label{fig:zspec_zphot_gpz}
\end{figure*}

In Fig.~\ref{fig:zspec_zphot_gpz} we present the results of our two \textsc{GPz} photo-$z$ estimates for the Bo\"{o}tes AGN in comparison to the consensus estimates produced through template-fitting in \citetalias{Duncan:2017wu}.
In each set of figures we show the distribution of photo-$z$ vs spectroscopic redshift for the consensus template estimates from \citetalias[][left]{Duncan:2017wu}, the \textsc{GPz} estimate with no weighting included in the cost-sensitive learning (centre) and the \textsc{GPz} estimate incorporating the colour and magnitude dependent weights as presented in Section~\ref{sec:weights} (right).
The sample plotted in each row contains only the subset of test sources not included in the training of the \textsc{GPz} classifiers.

To compare the quality of the different photo-$z$ estimates we make use of the same metrics as outlined in \citetalias[][we include the definitions in Table~\ref{tab:definitions}]{Duncan:2017wu}.
In Table~\ref{tab:photzstats} we present these photo-$z$ quality metrics for each of the AGN/galaxy subsamples.

\begin{table*}
	\centering
	\caption{Definitions of statistical metrics used to evaluate photometric redshift accuracy and quality along with notation used throughout the text.}\label{tab:definitions}

	\begin{tabular}{rll} 
		\hline
		Metric &  & Definition \\
		\hline
		$\sigma_{\text{NMAD}}$ & Normalised median absolute deviation & $1.48 \times \text{median} ( \left | \Delta z \right | / (1+z_{\text{spec}}))$ \\
		& Bias &$\text{median} (\Delta z )$\\
		O$_{f}$ & Outlier fraction & Outliers defined as $\left | \Delta z \right | / (1+z_{\text{spec}}) > 0.2$ \\
		$\overline{\textup{CRPS}}$ & Mean continuous ranked probability score &  $\overline{\textup{CRPS}} = \frac{1}{N} \sum_{i=1}^{N} \int_{-\infty}^{+\infty} [ \textup{CDF}_{i}(z) -  \textup{CDF}_{z_{s},i}(z)]^{2} dz$ - 
		 \citet{2000WtFor..15..559H} \\
		\hline
	\end{tabular}
\end{table*}

\begin{table}
\centering
\caption{Photometric redshift quality statistics for the derived combined consensus redshift predictions in the Bo\"{o}tes field. The statistical metrics (see Table~\ref{tab:definitions}) are shown for the full spectroscopic sample, the radio detected sources and for various subsets of the radio population. 
}\label{tab:photzstats}
\begin{tabular}{lccc}
Estimate & $\sigma_{\text{NMAD}}$ & Bias & O$_{f}$\\
\hline
\multicolumn{4}{c}{IR AGN}\\
Template consensus & 0.2429 &  0.0159 & 0.4425 \\
GPz - Unweighted & 0.1431 & -0.0187 & 0.2184 \\
GPz - Weighted & 0.1183 & -0.0072 & 0.1494 \\
\hline
\multicolumn{4}{c}{X-ray AGN}\\
Template consensus & 0.1067 & 0.0185	 & 0.3214 \\
GPz - Unweighted & 0.1241 & 0.0090 & 0.1339 \\
GPz - Weighted & 0.0882 & 0.0090 & 0.0893 \\
\hline
\multicolumn{4}{c}{Optical AGN}\\
Template consensus & 0.2351 & 0.0169	 & 0.4552 \\
GPz - Unweighted & 0.1280 & 0.0195 & 0.1970 \\
GPz - Weighted & 0.1147 & 0.0084 & 0.2313 \\
\hline
\multicolumn{4}{c}{Galaxies}\\
Template consensus & 0.0287 & -0.0037 & 0.0416 \\
GPz - Unweighted & 0.0323 & 0.0038 & 0.0220 \\
GPz - Weighted & 0.0343 & 0.0033 & 0.0265 \\
\end{tabular}
\end{table}

Visually, the poor performance of the template estimates for AGN populations between $1 \lesssim z \lesssim 3$ is clear in the left-hand column of Fig.~\ref{fig:zspec_zphot_gpz}.
Within this spectroscopic redshift range, many AGN sources are erroneously pushed towards $z\sim2$, albeit with large uncertainties that keep the photo-$z$ estimate within error of the true estimate.
Alternatively, sources at $1 \lesssim z \lesssim 3$ can have template estimates that are catastrophic failures, leading to estimated redshifts at $z \ll 1$.

Statistically, the overall improvement offered by the \textsc{GPz} estimates is illustrated in the reduction in scatter for the IR and optically selected AGN samples by a factor of two.
The improvement in scatter for the X-ray selected AGN subset is less drastic but still very significant - again most noticeably at $z > 1$.
As noted by \citet{Salvato:2008ef,Salvato:2011dq} many X-ray selected AGN are more accurately described by purely stellar SEDs - the template based photo-$z$s may therefore be expected to perform better for this subset than for the IR or optical AGN population.
Improvement in the measured outlier fractions is consistent across all three subsets, with the outlier fraction, O$_{f}$ (Table~\ref{tab:definitions}), measured for the \textsc{GPz} estimates typically a factor of two lower.


When applied to the remaining majority of galaxies that do not satisfy any of our AGN selection criteria, \textsc{GPz} is not able to significantly improve upon the estimates produced through template fitting -- at least not when restricted to using a set of filters that maximises the number of sources that can be fitted.
The performance of \textsc{GPz} with respect to the consensus template estimates is mixed, with $\approx 20\%$ worse scatter but $\approx 20-40\%$ better outlier fractions for the machine learning estimates.

\subsubsection{Accuracy of the error estimates}
Following Paper I and \citet{2016MNRAS.457.4005W}, we quantify the over- or under-confidence of our photometric redshift estimates by calculating the distribution of threshold credible intervals, $c$, where the spectroscopic redshift intersects the redshift posterior.
For a set of redshift posteriors which perfectly represent the redshift uncertainty, the expected distribution of $c$ values should be constant between 0 and 1, with the cumulative distribution $\hat{F}(c)$ therefore following a straight 1:1 relation as in a quantile-quantile plot (Q-Q).
Curves which fall below this expected 1:1 relation therefore indicate that there is overconfidence in the photometric redshift errors; the $P(z)$s are too sharp.

In the case of GPz, which provides only uni-modal Gaussian posterior redshift prediction with centre $z_{i,\textup{phot}}$ and width $\sigma_{i}$ (see Section~\ref{sec:method}), $c$ can be calculated for an individual galaxy analytically following
\begin{equation}
c_{i} = \Phi(n_{i}) - \Phi(-n_{i}) = \textup{erf} \left (\frac{n_{i}}{\sqrt{2}} \right),
\end{equation}
where $\Phi(n_{i})$ is the normal cumulative distribution function and $n_{i}$ can be simply calculated as $| z_{i,\textup{spec}} - z_{i,\textup{phot}} | / \sigma_{i}$.

For each \textsc{GPz} estimate we then implement the additional magnitude-dependent error calibration in a similar fashion to Paper I, varying the width of the Gaussian errors in order to minimise the Euclidean distance between the calculated distribution and the optimum 1:1 relation \citep[see also][for a similar analysis on uncertainty calibration for GPz estimates]{Gomes:2017ut}.
During the error calibration procedure, optimisation of the magnitude-dependent scaling parameters that minimise the Euclidean distance between observed and ideal distributions is done using only the test subsample (consisting of a random subset of 80\% of the total object).

\begin{figure*}
\centering
	\includegraphics[width=0.32\textwidth]{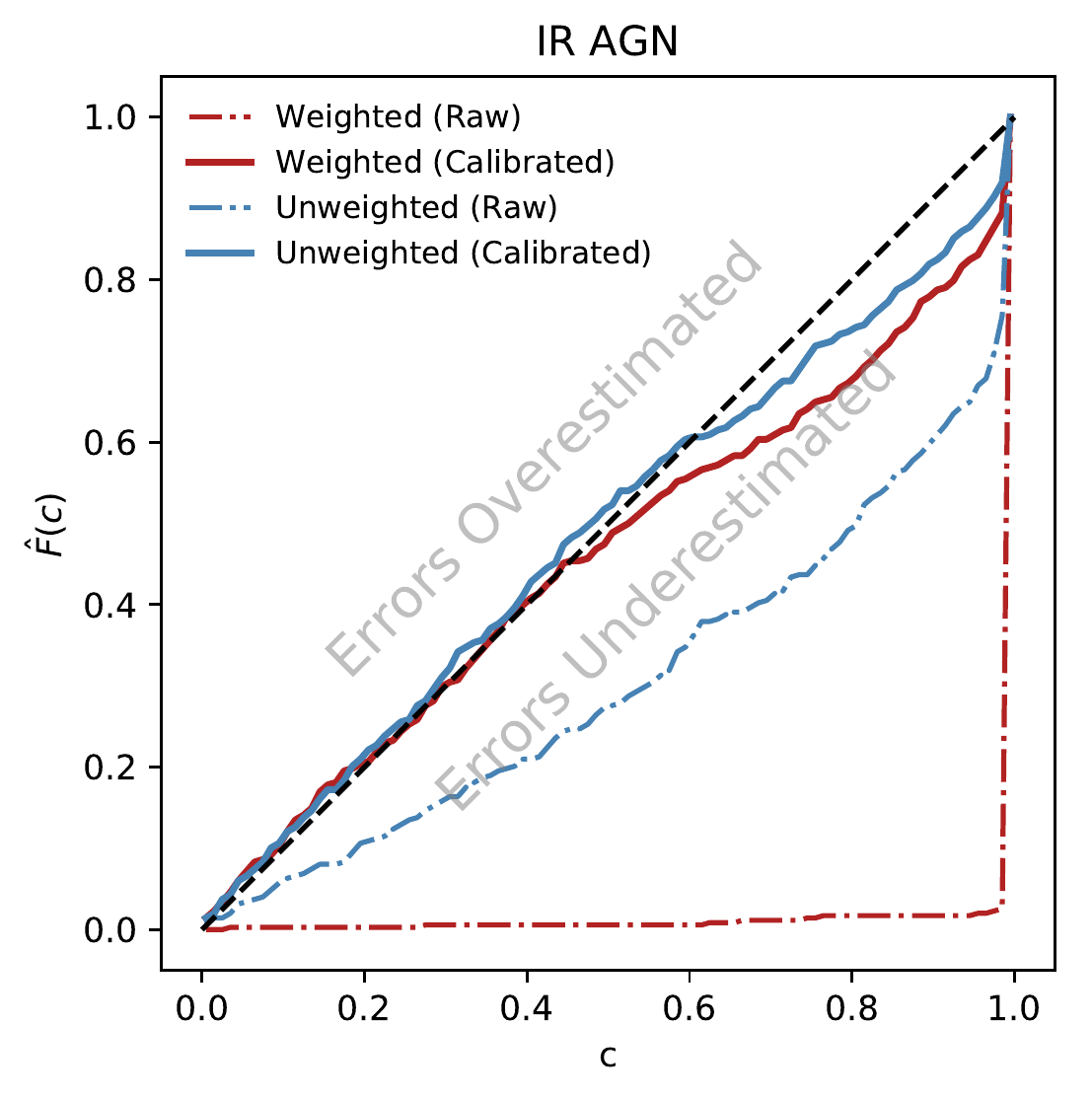} 
	\includegraphics[width=0.32\textwidth]{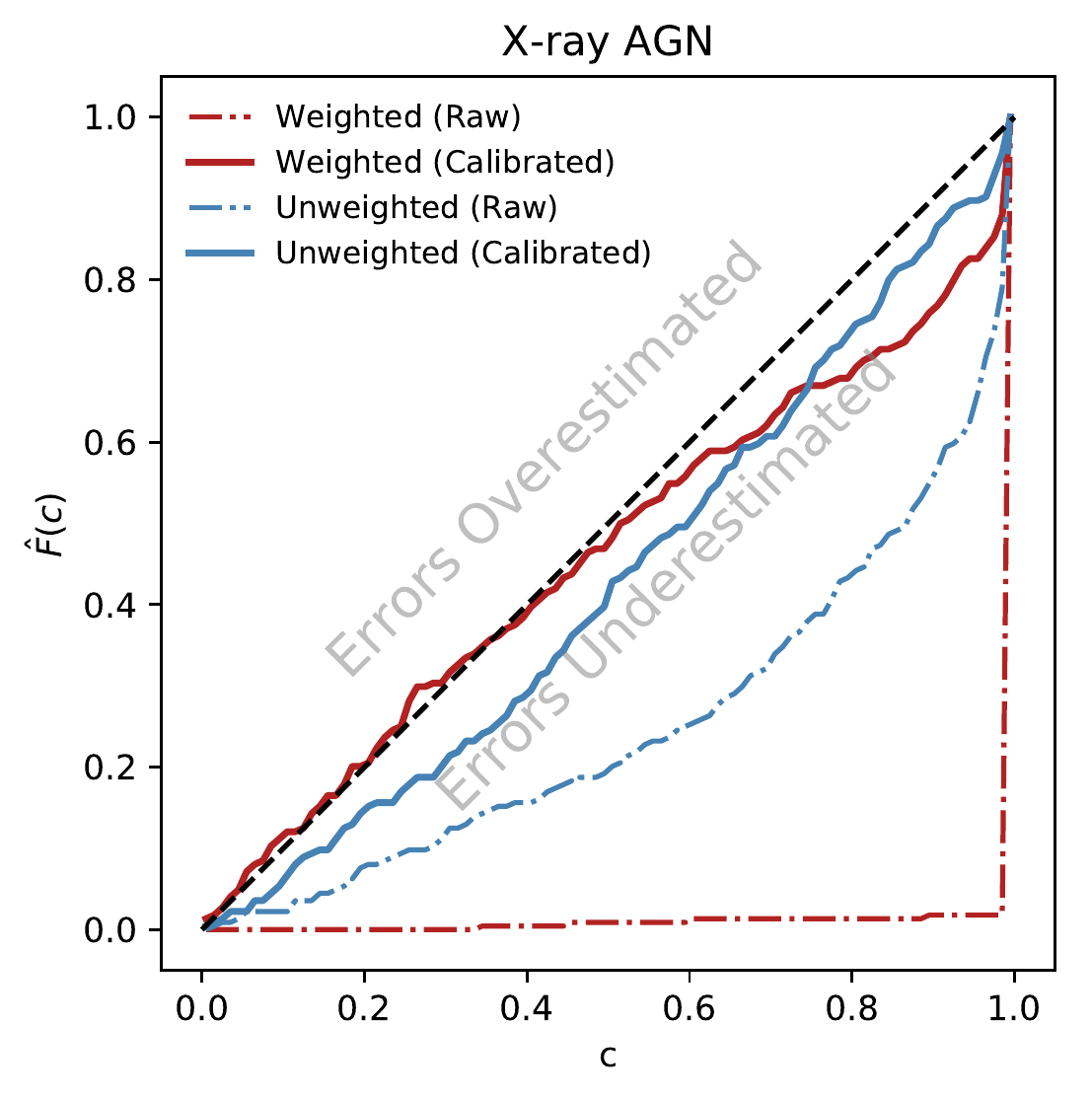} 
	\includegraphics[width=0.32\textwidth]{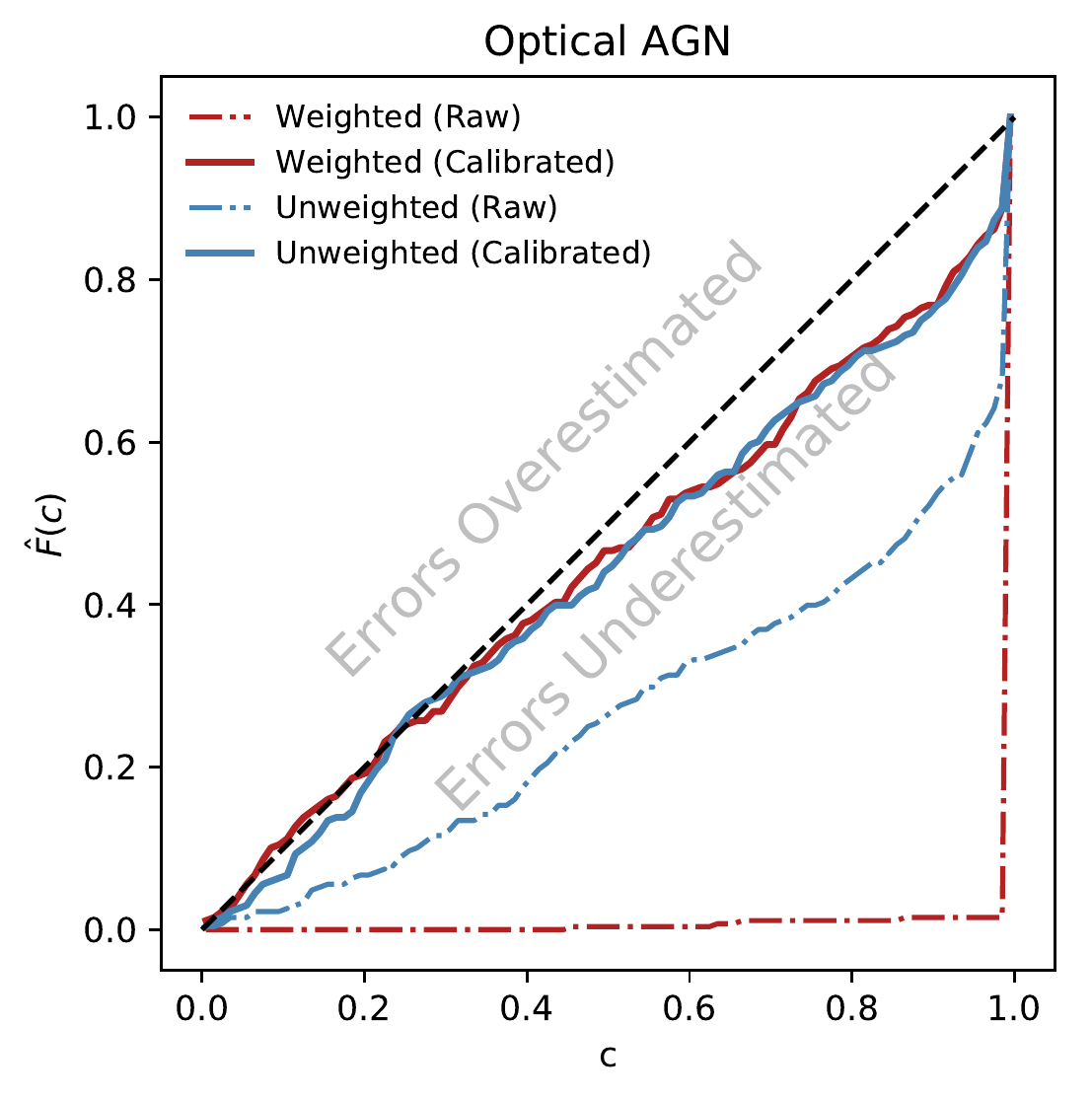} 

  \caption{Q-Q ($\hat{F}(c)$) plots for the redshift predictions for the two Gaussian process photo-$z$ estimates using unweighted (blue) and colour-magnitude weighted (red) Bo\"{o}tes training samples. The dot-dash and continuous lines show the results for the raw (as estimated by GPz) and calibrated distributions respectively. Lines that fall above the 1:1 relation illustrate underconfidence in the photo-$z$ uncertainties (uncertainties  overestimated) while lines under illustrate over-confidence (uncertainties underestimated).}
  \label{fig:qq_gpz}
\end{figure*}

In Fig.~\ref{fig:qq_gpz}, we present the Q-Q plots of the raw and calibrated error distributions for each of the three AGN estimators - plotting the results for the combined validation and test subsets (20\% of the complete subset) that were not included in the error calibration in any way.
Although \textsc{GPz} includes the accuracy of the uncertainties within the metric it aims to minimise, the redshift posteriors output still typically underestimate the photometric redshift uncertainty.
This overconfidence is consistent across all three AGN estimators but is noticeably worse when using the colour-magnitude weights in the cost-sensitive learning.
After the error calibration procedure has been applied, we see significant improvement in the accuracy of the redshift posteriors in almost all cases and errors that are close to the ideal solution.

\subsection{`Features' in the observed photometry}\label{sec:features}
The strong performance of the Gaussian process redshift estimates in the regime where those from template fitting struggle raises the question of what features in the optical/IR photometry is \textsc{GPz} using to derive the redshift information?
And secondly, are those features missing from the template sets employed in the previous photo-$z$ estimates?
Or is the failure due to other factors such as variability in the photometry?

\begin{figure}
\centering
	\includegraphics[width=0.98\columnwidth]{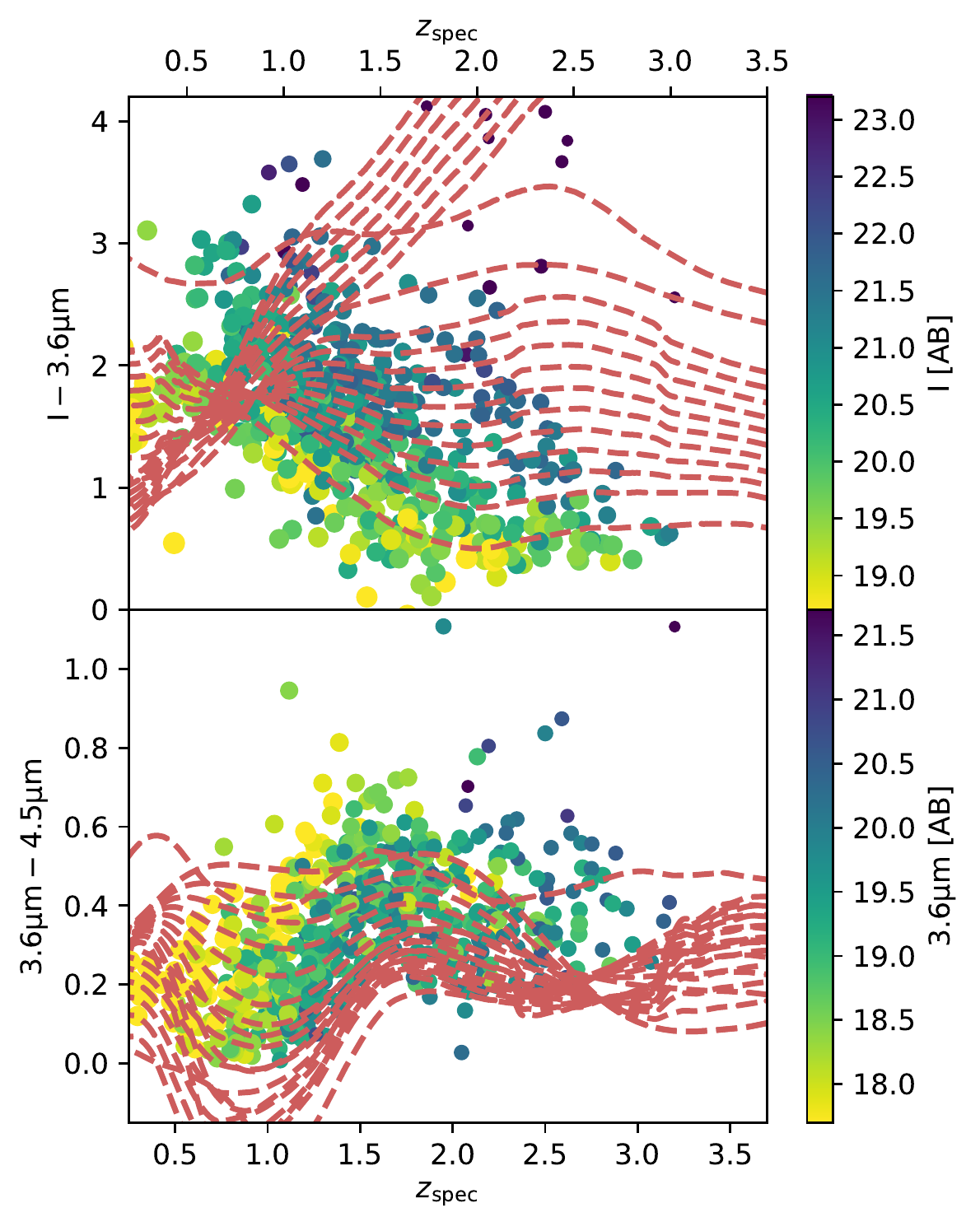} 

  \caption{Selected observed colours as a function of redshift for the Bo\"{o}tes IR-selected AGN population. The upper panel shows the optical to mid-IR colour between the $I$ and IRAC $3.6\mu \textup{m}$ bands while the lower panel shows the mid-IR colour between the IRAC $3.6\mu \textup{m}$ and $4.5\mu \textup{m}$ bands. In each panel, the colour of the datapoints corresponds to the apparent magnitude in one of the observed bands. Dashed red lines indicate the colour-tracks as a function of redshift for the XMM-COSMOS \citep{Salvato:2011dq} templates which satisfy the IR AGN selection criteria of \citet{Donley:2012ji} at any redshift up to $z=3$.}
  \label{fig:iragn_colour_relations}
\end{figure}

Investigating the cause of each template-based photo-$z$ failure individually is beyond the scope of this paper.
However, we can very easily verify the existence of redshift-dependent colour or magnitude relations upon which the empirical photo-$z$s might be deriving their results. 
To illustrate this, in Fig.~\ref{fig:iragn_colour_relations} we show how two example colours and corresponding apparent magnitudes evolve with redshift for the IR selected AGN population.
In the redshift regime of $1 < z < 2$ where \textsc{GPz} performs exceedingly well, it is clear that there is a strong evolution in the $3.6\mu \textup{m} - 4.5\mu  \textup{m}$ colour (with a strong feature at $z\sim 1.7$) while the typical $I - 3.6\mu \textup{m}$ also become increasingly blue over this range. 
Coupled with the colour-redshift relations are complementary magnitude-redshift relations for the optical and mid-IR bands - the evolution of $I$-band magnitude for a fixed $I - 3.6\mu \textup{m}$ colour with redshift at $z \gtrsim 1$ remains relatively constant while the apparent $3.6\mu \textup{m}$ magnitude shows a much clearer trend of fainter magnitudes at higher redshift.
Altogether it is therefore clear that at least for the IR AGN population, there are redshift dependent magnitude or colour features to which we can anchor empirical photo-$z$ estimates.

The follow-up question raised at the beginning of this section was whether the features \textsc{GPz} is basing its redshift predictions from are absent within the templates.
Sticking with the example of IR AGN, the bump in $3.6\mu \textup{m} - 4.5\mu  \textup{m}$ at $z\sim1.7$ is not well represented in the \citet{Brown:2014jd} library - which does not include powerful AGN.
But as illustrated by the colour tracks in Fig.~\ref{fig:iragn_colour_relations}, the \citet[][see also \citet{Hsu:eu}]{Salvato:2011dq} template set is able to fill the broad colour region of interest at most redshifts.

There are areas within the colour inhabited by the IR-selected AGN population that the templates do not cover, specifically they do not extend to blue enough $I - 3.6\mu \textup{m}$ colours at $z > 1$ and at $3.6\mu \textup{m} - 4.5\mu  \textup{m}$ the templates are no longer representative for this population in this colour-space.
Nevertheless, these deficiencies alone are unlikely to account for the very poor template performance at $z < 2$ and there may be an additional root causes for these failures.
Examination of the average residuals measured for the best-fit templates (both for the free redshift determination and when the redshift is fixed to the known spectroscopic redshift) find no clear indication that any one individual band or colour is responsible for the causing incorrect fits.

Future extensions to the existing template libraries that better sample the full AGN colour space (Brown et al. in preparation) will still likely offer significant improvements in this regime.
Furthermore, imposing a strong mid-infrared magnitude prior specific to the source-type may aid the template based estimates by breaking degeneracies in colour space (e.g. see the lower panel of Fig.~\ref{fig:iragn_colour_relations}).
Due to the focus of this study on the \textsc{GPz} estimates, we defer any further investigation of the AGN template properties to future studies and instead concentrate the rest of our analysis on the machine learning estimates and those derived from them.

\section{`Hybrid' photo-$z$s - Combining GP redshift estimates with template estimates}\label{sec:combined}

\begin{figure*}
\centering
	\includegraphics[width=0.98\textwidth]{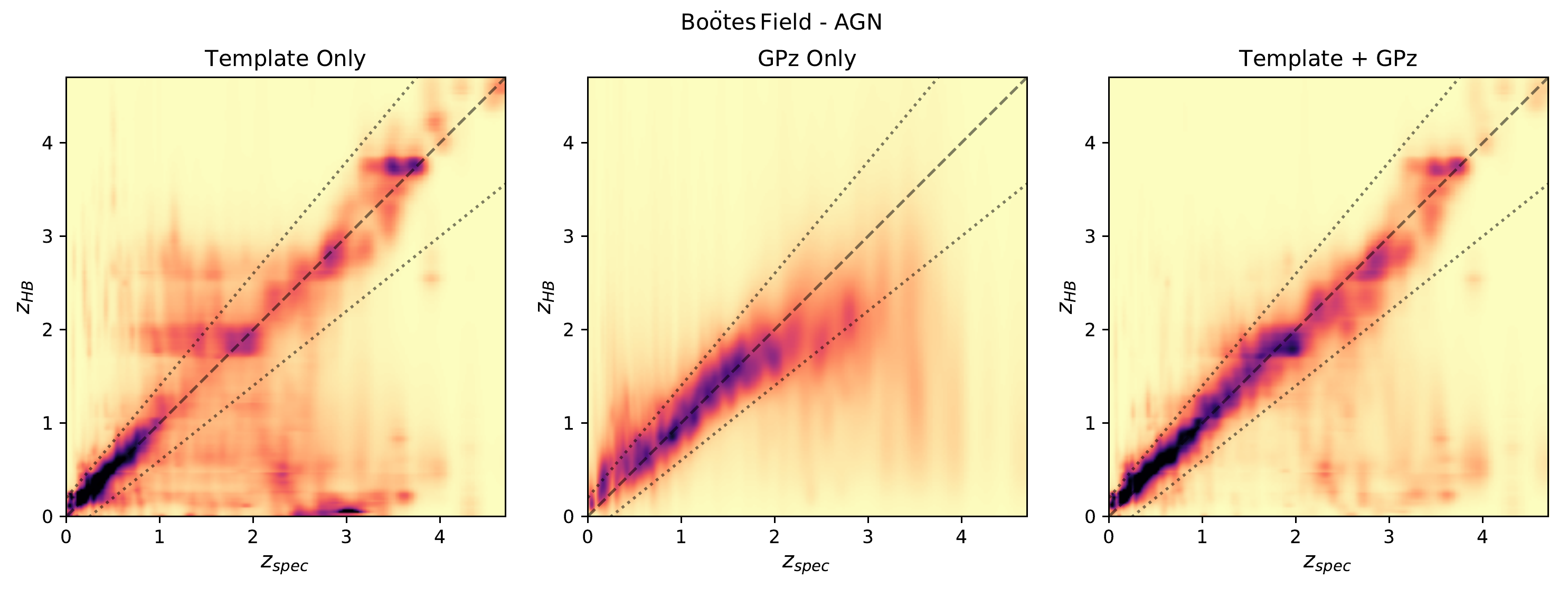} 

  \caption{Stacked probability distributions for the combined AGN population (IR, X-ray or optically selected) as a function of spectroscopic redshift for each consensus HB photo-$z$ estimate. To improve the visual clarity at higher redshifts where there are few sources within a given spectroscopic redshift bin, the distributions have been smoothed along the x-axis. The same smoothing has been applied to all three estimates consistently. The dashed grey line corresponds to the 1:1 relation while the dotted lines correspond to the outlier definition adopted in this work. The superior performance of the hybrid template + \textsc{GPz} estimates is well illustrated by the side-by-side comparison.}
  \label{fig:hb_pz_comparison}
\end{figure*}

One of the key conclusions of \citetalias{Duncan:2017wu} and earlier studies in the literature \citep[e.g.][]{Dahlen:2013eu,CarrascoKind:2014jg} was that no single photometric estimate can perform the best for all source types or in all metrics.
Furthermore, the combination of multiple estimates within a statistically motivated framework can yield consensus estimates that perform better than any of the individual inputs.
Given the very different limitations and systematics observed in the template and \textsc{GPz} photo-$z$ estimates, a consensus photo-$z$ that compounds the advantages of both methods is clearly desirable.

To incorporate the \textsc{GPz} predictions within the hierarchical Bayesian (HB) combination framework presented in \citetalias{Duncan:2017wu}, normal distributions based on position and corrected variance estimate for each source are evaluated onto the same redshift grid as used during the template fitting procedure.
For any source in the full training sample that does not have a photo-$z$ estimate for a given \textsc{GPz} estimator (either through not satisfying the selection criteria for a given subset or lack of observations in a required band), we assume a flat redshift posterior over the range of the redshift grid ($P(z) = 1/7$).
These sources therefore contribute no information in the HB combination procedure, so in the cases where only one estimate exists the consensus estimate is entirely based on that single prediction.

For comparison with the template-based consensus estimates from \citetalias{Duncan:2017wu}, we calculate two different HB estimates from our \textsc{GPz} estimates.
Firstly, we calculate the HB consensus photo-$z$ based only on the four separate \textsc{GPz} estimates (optical, X-ray and IR AGN estimates plus the additional galaxy-only estimate).
Secondly, we then calculate the HB consensus estimate incorporating all three of the template based estimates calculated in \citetalias{Duncan:2017wu} and the four machine learning estimates from this paper to produce a hybrid estimate.
In both cases we follow the practice of \citetalias{Duncan:2017wu} and adopt a magnitude based prior when an observation is assumed to be `bad'.

In Fig.~\ref{fig:hb_pz_comparison} we present the photo-$z$ vs spectroscopic redshift distribution of the three separate HB consensus estimates.
To better illustrate the overall uncertainty and scatter given the large number of sources, we show the stacked redshift probability distributions within a spectroscopic redshift bin rather than individual point estimates.
The left panel of Fig.~\ref{fig:hb_pz_comparison} illustrates the previously known limitations of template-based photo-$z$ estimates for most AGN sources.
At $z < 1$ the template estimates perform well, but between $1 < z < 3$ the photo-$z$ probability distributions are extremely broad; possibly due to the lack of strong photometric features in the optical SEDs in this regime.
Additionally, the degradation of the template photo-$z$ quality towards higher redshift may be a result of differences in the source population selected at higher redshift; the galaxy templates work well for the low-luminosity AGN but fail for higher luminosity AGN where the host galaxy no longer dominates the optical emission.
At $z \gtrsim 3$, the template-based estimates begin to perform well again due to the redshifted Lyman-continuum break moving into the observed optical bands. 

It is worth noting that the extent of the template photo-$z$ issues at $1 < z < 3$ are partly field specific, in that the relative depths of the near-infrared data available in the Bo\"{o}tes field are shallow with respect to the optical and mid-infrared data at wavelengths either side. 
As such, sources which may have high signal-to-noise ($\textrm{S/N}$) detections in the optical regime may still have very low $\textrm{S/N}$ in the near-IR bands that probe the rest-frame optical features (both in spectral breaks and emission lines) at $z \gtrsim 1$.
Fig. 5 of \citetalias{Duncan:2017wu} shows that in fields with deeper photometry and finer wavelength coverage \citep[e.g. the COSMOS field][]{Laigle:2016ku} the trends are not as extreme, particularly at $1 < z < 2$. 
Nevertheless, the improvement seen here is particularly encouraging for photo-$z$ estimates in surveys without the same levels of exceptional filter coverage as available in COSMOS.

In contrast to the trends observed in our template estimates, and consistent with the trends seen in individual AGN estimates shown in Fig.~\ref{fig:hb_pz_comparison}, the GPz-only consensus estimates perform best in the region of $1 \lesssim z \lesssim 2$.
At lower ($z \lesssim 0.5$) and higher ($z \gtrsim 2.5$) redshifts, the \textsc{GPz} consensus estimate becomes increasingly biased.
It is these wavelength regimes in which the training samples for the AGN population are most sparse, as can be seen visually in the right hand column of Fig.\ref{fig:zspec_zphot_gpz}.

Most encouraging however is the HB consensus estimate incorporating both the template and machine learning based predictions (right panel of Fig.~\ref{fig:hb_pz_comparison}).
Visually, it is immediately clear that the total combined consensus estimate combines the advantages of both of the input methods.

\begin{figure}
\centering
\includegraphics[width=0.95\columnwidth]{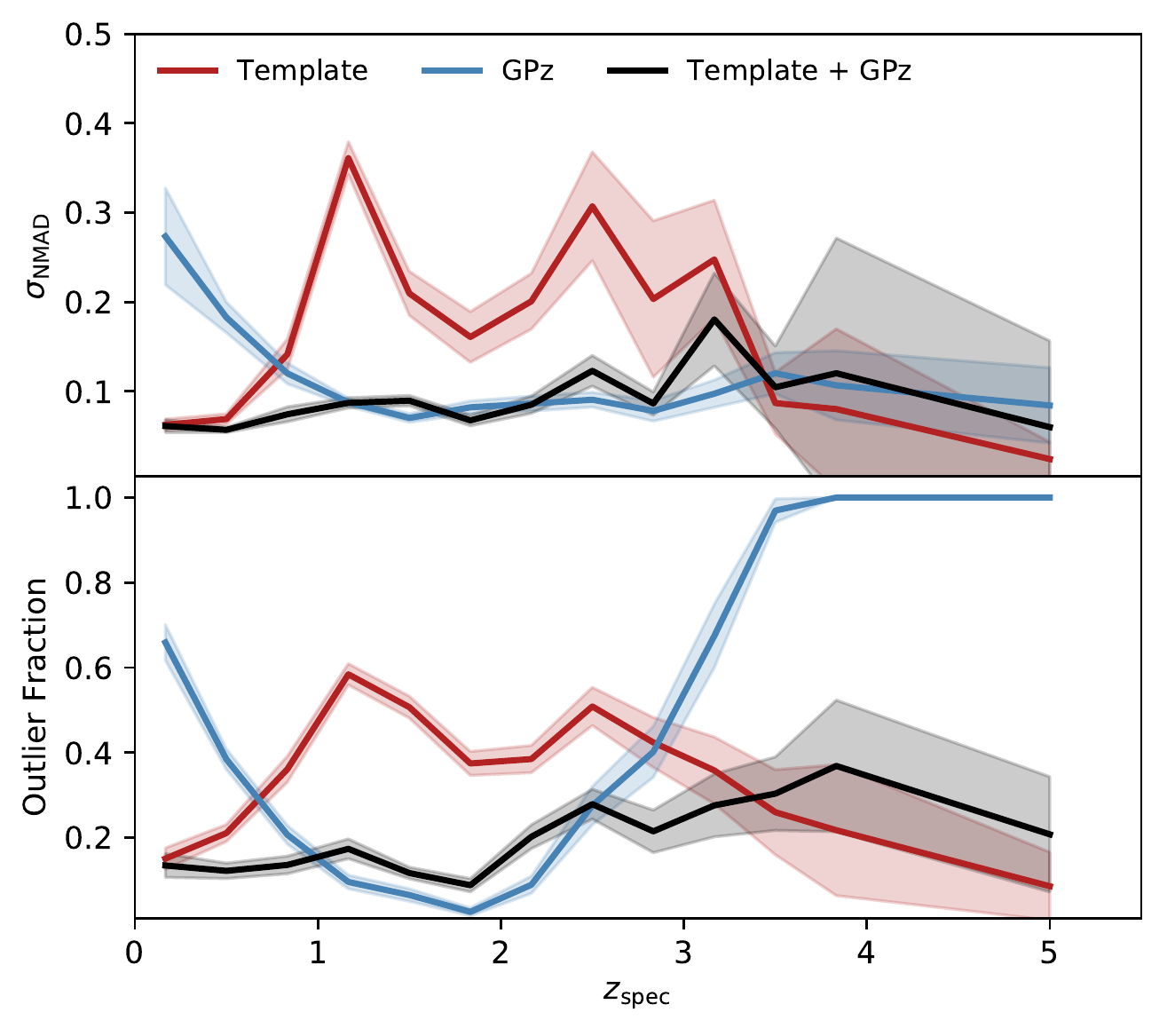}
  \caption{Photometric redshift scatter ($\sigma_{\text{NMAD}}$) and outlier fraction as a function of spectroscopic redshift for AGN in the Bo\"{o}tes field. Lines show the results for sources that pass any of the X-ray/Optical/IR AGN criteria outlined in Section~\ref{sec:data}. Shaded regions around each line represent the standard deviation on the corresponding metric from Bootstrap resampling.}
  \label{fig:sigma_vs_z}
\end{figure}

\begin{figure}
\centering
\includegraphics[width=0.95\columnwidth]{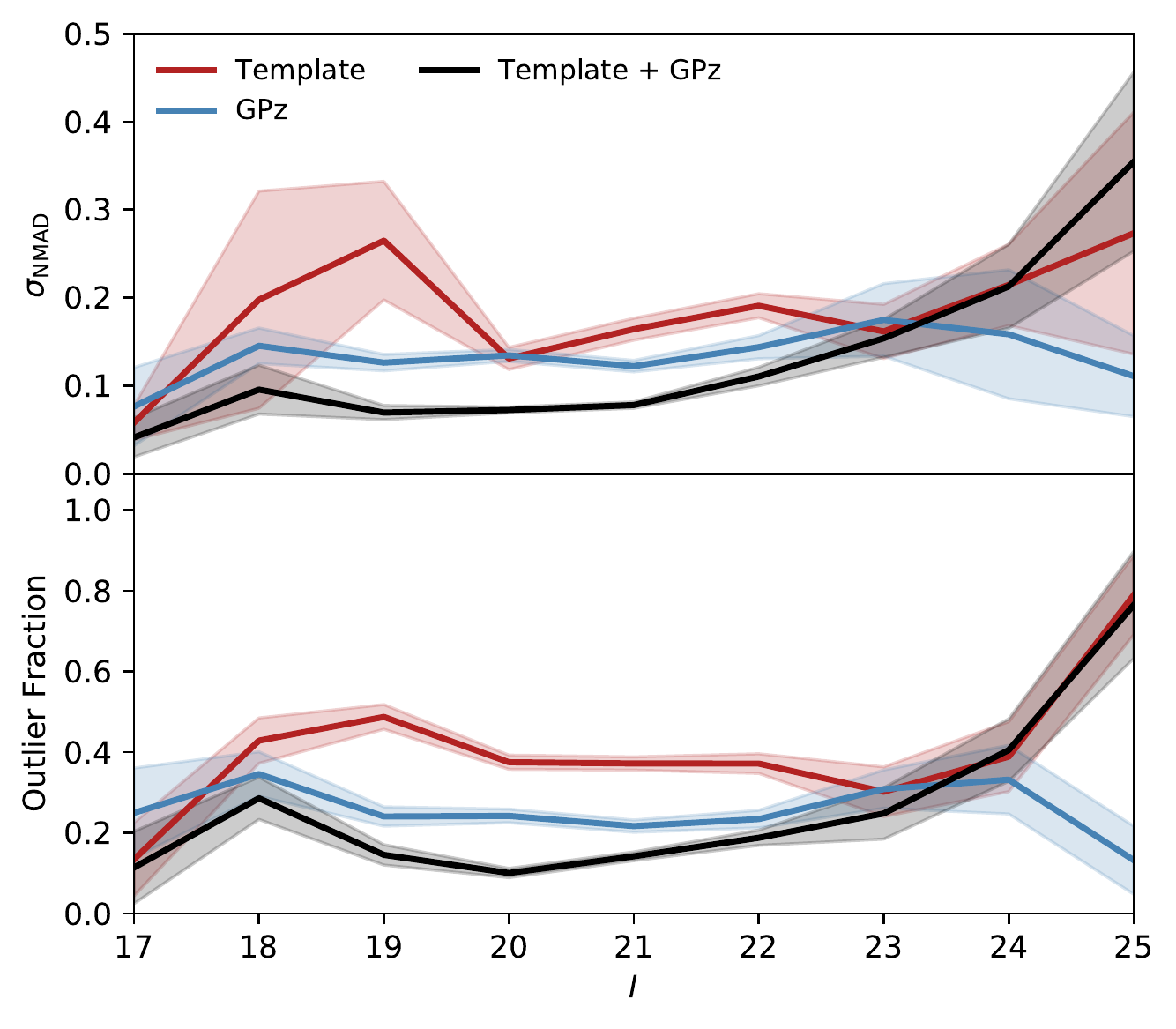}
  \caption{Photometric redshift scatter ($\sigma_{\text{NMAD}}$) and outlier fraction as a function of $I$ magnitude for AGN sources in the Bo\"{o}tes field. Lines show the results for sources that pass any of the X-ray/Optical/IR AGN criteria outlined in Section~\ref{sec:data}. Shaded regions around each line represent the standard deviation on the corresponding metric from Bootstrap resampling. At almost all redshift ranges, the hybrid photo-$z$ performance is comparable or better to the best input methodology.}
  \label{fig:sigma_vs_mag}
\end{figure}

This improvement can also be seen more quantitatively by looking at the measured photo-$z$ scatter and outlier fraction for the AGN population as a function of redshift (Fig.~\ref{fig:sigma_vs_z}).
At $z < 1$, the hybrid estimates match or improve upon the scatter from the template estimates.
Then, at $1 < z < 3$, the hybrid estimates match the improved scatter and outlier fractions of the \textsc{GPz} estimates while the template-based estimates perform very poorly.
Finally, at $z \gtrsim 3$ when strong continuum features result in improved template estimates, the hybrid estimates are still able to perform comparably.

Fig.~\ref{fig:sigma_vs_mag} shows the measured scatter and outlier fraction as a function of apparent $I$-band magnitude. 
At all magnitudes brighter than $I \approx 23.5$, the hybrid estimates perform better than either the template or \textsc{GPz} only estimates.
The observed improvement in scatter for the \textsc{GPz}-only estimates at the very faintest magnitudes (as compared to the template or hybrid method) likely results from the cost-sensitive learning increasing the importance of these faint AGN during the optimisation procedure.
However, it is evident that the hybrid estimates are most similar in performance to the template-only estimates in this regime, with the rise in scatter and outlier fraction at $I > 23$ closely mirroring the observed rise. 
The apparent inability of the hybrid consensus estimates to mirror the performance of the best performing estimate could be seen as a failure of the hierarchical Bayesian combination method at faint magnitudes.

\begin{figure}
\centering
  \includegraphics[width=0.99\columnwidth]{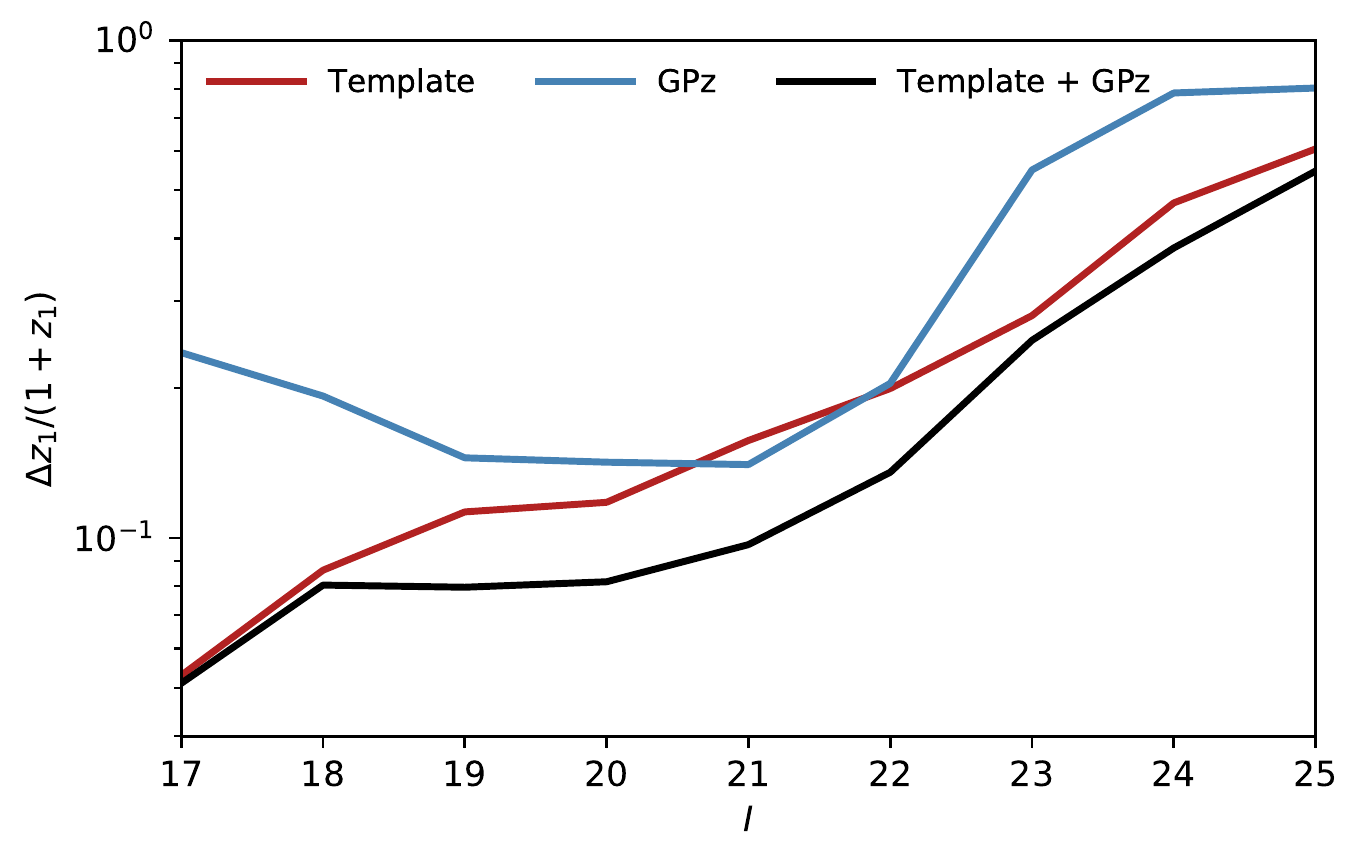}
  \caption{Average difference between the median of the primary redshift peak, $z_{1}$, and upper 80\% highest probability density credible interval, denoted here as $\Delta_{z_{1}}$, in bins of apparent $I$-band magnitude for the AGN sources in the Bo\"{o}tes field. We illustrate only the upper error bounds to improve clarity by allowing a logarithmic scale. Within the primary peak, positive and negative errors are found to be very symmetrical; negative errors for each estimate follow the same magnitude trends.}
  \label{fig:sigma_err_vs_z}
\end{figure}

For all three estimates, the posterior distributions are very broad at faint magnitudes.
Evidence for this can be seen in Fig.~\ref{fig:sigma_err_vs_z}, where we show the median difference between the median of the primary redshift peak and the upper 80\% highest probability density (HPD) credible interval as a function of magnitude for the three consensus estimates.

Visual inspection of the three different consensus redshift posteriors for the very faintest sources ($I > 24$) therefore reveals that for the \textsc{GPz}-only consensus estimate, the uncertainties on the individual estimates are so large at the faint magnitudes that the consensus $P(z)$ is dominated by the redshift prior.
The apparent improvement in the accuracy of the \textsc{GPz} redshift estimates is therefore something of a conspiracy - with the median redshift of the redshift prior (for $I > 24$) lying close to the average spectroscopic redshift for these sources.

We note here that this magnitude regime is at the limits of the Bo\"{o}tes optical data; beyond $I \sim 24$ the typical source S/N becomes very low and the catalogs increasingly incomplete.
As such, we do not expect photo-$z$ performance to remain good enough at these magnitudes for most scientific purposes.

\subsection{Comparison to \citet{Brodwin:2006dp}}
As mentioned in the introduction, this study is not the first to attempt to combine the different strengths of template-based and empirical photo-$z$ estimates.
In addition to the comparison of different methods for Bayesian combination of template and machine learning estimates presented in \citet{CarrascoKind:2014jg},  \citet{Brodwin:2006dp} have also previously explored a hybrid photo-$z$ method aimed at improving estimates for AGN within the Bo\"{o}tes field.

Based on predominantly the same underlying photometry as used in this analysis, \citet{Brodwin:2006dp} estimated photo-$z$s using two approaches - firstly using template fitting and secondly employing an empirical method using neural networks \citep{Collister:2004fx}.
The most direct comparison we are able to make between the results of \citet{Brodwin:2006dp} and those presented in this work is via their quoted estimates of the 95$\%$-clipped photo-$z$ scatter.

For AGN between $0 < z < 3$ in the AGES \citep{Kochanek:jy} spectroscopic sample, \citet{Brodwin:2006dp} find a scatter of $\sigma_{95\%}/(1+z) = 0.12$ and for galaxies between $0 < z < 1.5$ a lower scatter of $\sigma_{95\%}/(1+z) = 0.047$.
Restricting our spectroscopic sample to contain only those from AGES and requiring a $4.5\mu m$ detection to best match the \citeauthor{Brodwin:2006dp} selection criteria, our hybrid photo-$z$ estimate have comparable 95$\%$-clipped scatters of $\sigma_{95\%}/(1+z) = 0.11$ and $\sigma_{95\%}/(1+z) = 0.045$ for sources classified by AGES as AGN and galaxies respectively.

When comparing the two results it is important to recognise that the template-fitting and the GPz estimates trained for the galaxy population make use of additional photometry not available at the time of \citet[][e.g. $u$, $z$ and $y$]{Brodwin:2006dp}.
Some small improvement is therefore to be expected.

A key improvement offered by the Bayesian combination framework employed in this work is that it is able to make maximal use of the redshift information available for a given source.
In \citet{Brodwin:2006dp}, the choice of template or neural-network based estimates for a given source is a binary based on where a source lies with respect to the \citet{Stern:2005ey} IRAC colour criteria (similar to the criteria we have used for selecting IR AGN).
As seen in Fig.~\ref{fig:zspec_zphot_gpz} the performance of machine learning estimates for these sources is significantly better over the redshift range of interest, so this choice is well motivated.
However, at higher redshifts the machine learning estimates become increasingly biased due to the sparsity of the training samples in this regime.
This bias is clearly visible both in Fig.~5 of \citet{Brodwin:2006dp} and in the centre panel of Fig.~\ref{fig:hb_pz_comparison} of this work.
Although still imperfect, the hierarchical Bayesian combination procedure is able to fall back on the more accurate and reliable template-based estimates at $z \gtrsim 2.5$.

\subsection{Hybrid photo-$z$ performance for the radio source population}
\begin{figure}
 \centering
  \includegraphics[width=0.96\columnwidth]{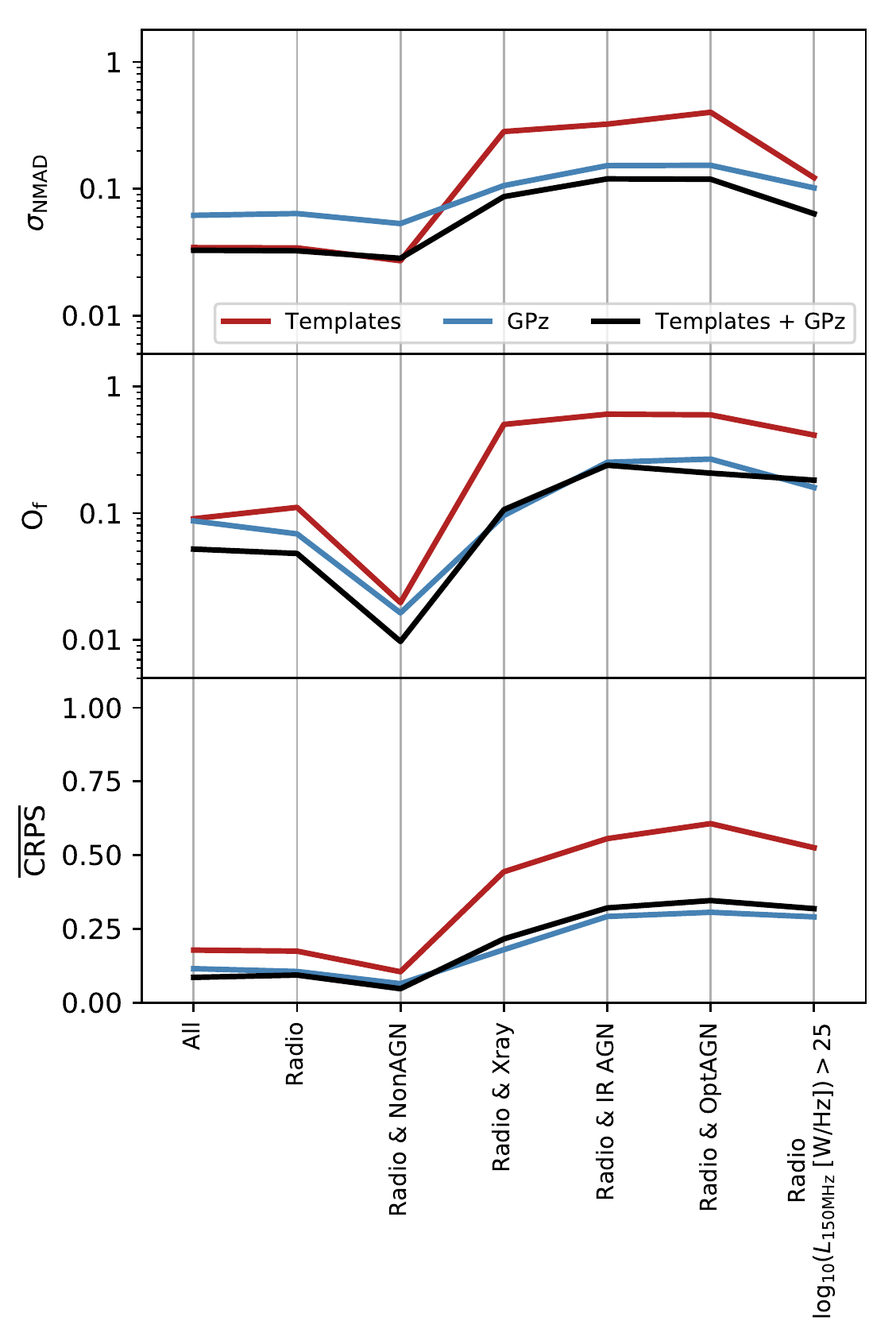}
  \caption{Visualised photometric redshift performance in three metrics ($\sigma_{\textup{NMAD}}$, $O_{\textup{f}}$, $\overline{\textup{CRPS}}$; see Table~\ref{tab:definitions}) for the different Bo\"{o}tes field radio source subsamples. For all subsets of the radio-detected population, the hybrid method performs better than either template or \textsc{GPz} alone.}
  \label{fig:lin_stats_bootes}
\end{figure}

Given our motivation in producing the best possible photo-$z$ estimates for the diverse population selected objects in forthcoming radio continuum surveys, it is interesting to see how the improvement seen in the optical/IR/X-ray selected AGN population propagates through into the hybrid photo-$z$ performance for radio selected objects.
In Fig.~\ref{fig:lin_stats_bootes} we illustrate the $\sigma_{\textup{NMAD}}$, $O_{\textup{f}}$ and $\overline{\textup{CRPS}}$\footnote{In \citetalias{Duncan:2017wu} we introduced the mean continuous ranked probability score, $\overline{\textup{CRPS}}$ as a performance metric that measures not just the accuracy of the photometric redshifts but also their relative precision. As with the scatter and outlier fraction, values as low as possible are desired.} performance of the template, \textsc{GPz} and hybrid consensus redshift estimates in each of the source population subsets. 
Across all subsets of the radio detected populations, the hybrid photo-$z$ estimates either match or significantly improve upon the scatter and outlier fraction performance of the best single method.

Furthermore, across all subsets of the radio population the scatter is now $\sigma_{\textup{NMAD}} \lesssim 0.1$, an improvement of up to a factor of four compared to the template estimates.
Despite them not performing significantly better than the template estimates for sources not optically classified as AGN, the inclusion of \textsc{GPz} estimates in the hierarchical Bayesian photo-$z$s results in a factor of $\sim2$ improvement in outlier fraction for the radio-detected subset of these sources.

\begin{figure}
 \centering
  \includegraphics[width=0.96\columnwidth]{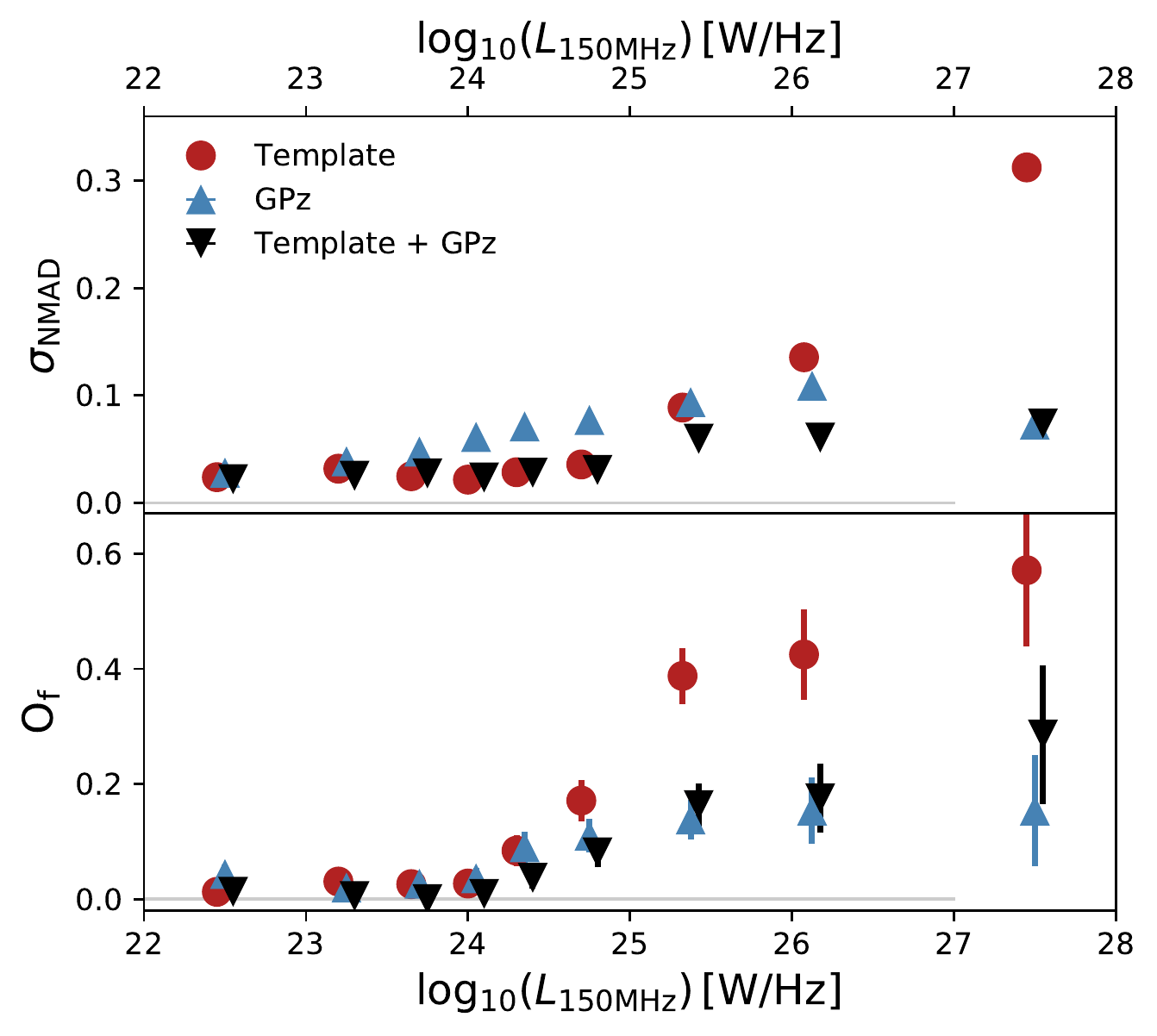}
  \caption{Photometric redshift scatter ($\sigma_{\textup{NMAD}}$; upper panel) and outlier fraction ($O_{f}$; lower panel) as a function of 150MHz radio luminosity for all radio detected Bo\"{o}tes field sources within the spectroscopic redshift range $0 < z < 3$. In each plot we show the values for the template-only (circles), GPz-only (upward triangles) and combined (downward triangles) consensus estimates. Symbols have been offset horizontally only for clarity, luminosity bins for all estimates are identical. Error-bars plotted for the outlier fractions illustrate the binomial uncertainties on each fraction. The hybrid estimate performs significantly better than either the template or GPz-only estimates across the full range of radio luminosities probed in this field, with particularly large improvement at the greatest radio powers.}
  \label{fig:stats_lum}
\end{figure}

\begin{figure}
 \centering
  \includegraphics[width=0.96\columnwidth]{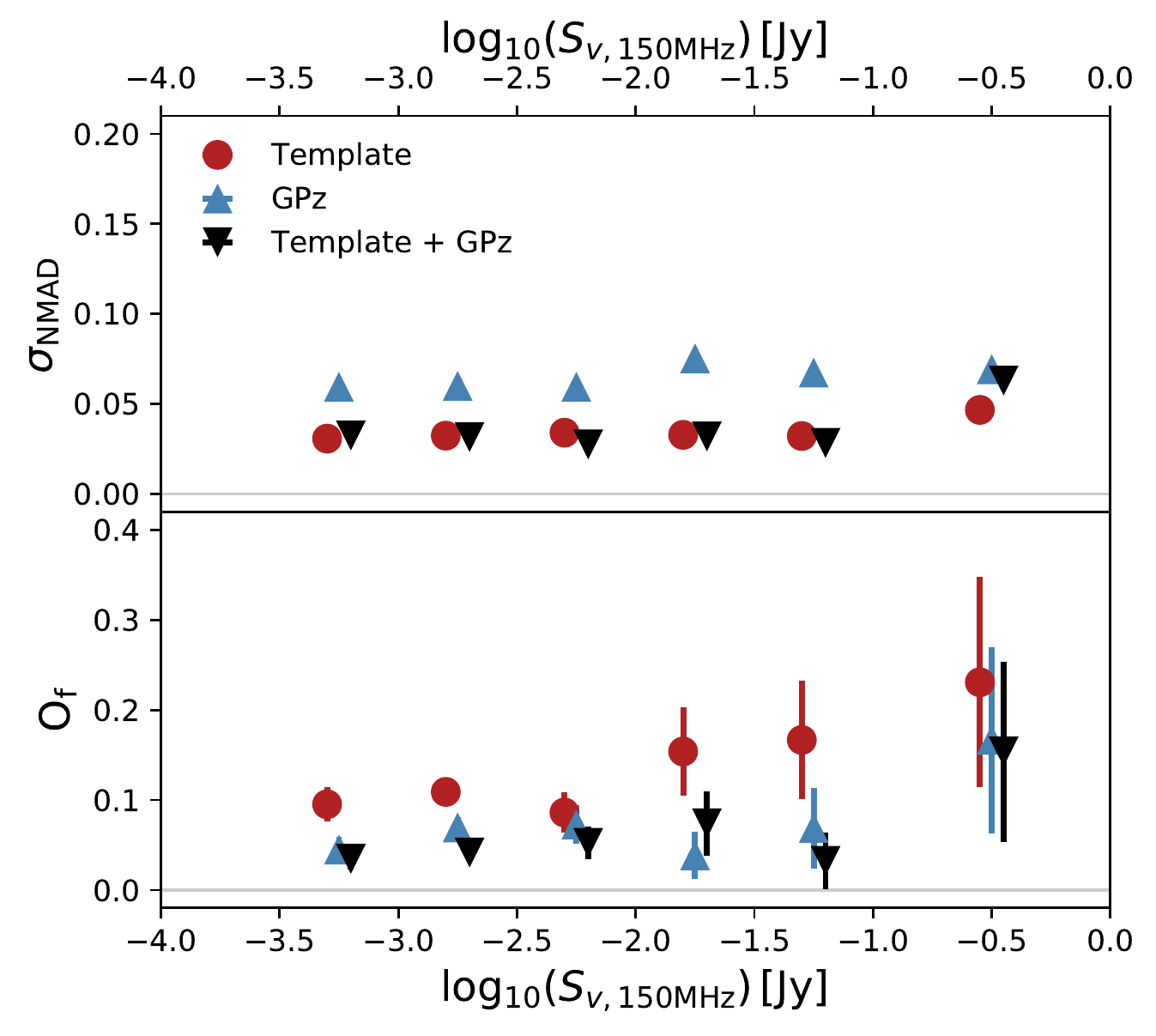}
  \caption{As in Fig.~\ref{fig:stats_lum} but for 150MHz radio flux density - for all radio detected sources within the spectroscopic redshift range $0 < z < 3$. Due to the majority of the spectroscopic training sample probing low redshift sources where the template estimates perform well, the improvement in the scatter for the hybrid estimates is not significant. However, the number of catastrophic outliers in the hybrid estimates is lower than the template-only estimates at all fluxes.}
  \label{fig:stats_flux}
\end{figure}

Exploring the key quality statistics as a function of radio luminosity (Fig.~\ref{fig:stats_lum}) and flux density (Fig.~\ref{fig:stats_flux}) we can see more clearly that the greatest gain in improvement is for the most luminous radio sources.
For a given apparent radio flux, the \textsc{GPz} and hybrid estimates offer no clear improvement in terms of scatter but do improve the outlier fraction.
This behaviour is something we would expect to see, bearing in mind that lower luminosity sources at low redshift dominate the spectroscopic sample we are comparing ($\approx 90\%$ of the spectroscopic sample is at $z < 1$).
The rarer high luminosity radio sources for which \textsc{GPz} produces more accurate photo-$z$ estimates have a broad range of apparent fluxes and therefore the robust scatter is not strongly affected but the outlier fraction is.

The performance of the \textsc{GPz}-only estimates compared to the template-only estimates as a function of radio power could shed further light on the discussion in Section~\ref{sec:features} on the causes of failures in the template fitting.
That \textsc{GPz} performs best for the most luminous radio AGN could support the idea that our selected template fits struggle most in the regime where the AGN dominates the optical emission.
Although in the local Universe the most powerful radio sources are typically host-dominated in their optical emission, at higher redshifts the population the population of QSO/Seyfert-like sources becomes increasingly dominant \citep[e.g.][and references therein]{2014ARA&A..52..589H,2018MNRAS.475.3429W}.
Within a deep survey field such as that used in this work, the larger volume probed at high redshift means that $z > 1$ sources dominate the high-luminosity end of our sample.
Further exploration of the different methods as a more detailed function of radio luminosity and redshift would clearly be valuable in better understanding our methods and their strengths and limitations, however the currently limited training sample makes this impractical. 

\subsection{Performance in deep optical fields}\label{sec:deep}
As outlined in Section~\ref{sec:data}, the decision to concentrate the analysis in this paper on the shallower and wider Bo\"{o}tes data was motivated partly by the more limited training sample available for AGN in the COSMOS field -- a field with significantly deeper and more extensive optical data.
Nevertheless, we apply the hybrid methodology to the COSMOS sample to explore how the hybrid method performs in a regime where template photo-$z$s generally perform exceptionally, i.e. with extensive deep photometric datasets that include fine sampling over optical wavelengths (through medium band photometry in the case of COSMOS).
Since we claim an advantage of the hybrid method is that it should optimally combine the information from different estimates, we would therefore expect the method to also be able to cope with the combination of more precise template estimates with potentially poorer machine learning estimates - while still incorporating any additional information they provide. 

We apply the \textup{GPz} method to the COSMOS dataset in the same way as for Bo\"{o}tes, with \textup{GPz} trained on subsets of the IR, X-ray and optical AGN population as well as the main galaxy population \citepalias[see ][for details on the AGN classifications used for COMSOS]{Duncan:2017wu}.

The bands chosen for each subset and the total number of training sources available for those bands (with 80\% used for training, 10\% for validation and 10\% for testing) are as follows:
\begin{itemize}
\item	 IR AGN: $r$, $i^{+}$ $z^{++}$, $3.6\mu$m, $4.5\mu$m, $3.6\mu$m and $4.5\mu$m bands (325 training sources)
\item X-ray AGN: $b$, $r$, $i^{+}$, $z^{++}$ ,$3.6\mu$m and $4.5\mu$m (1488)

\item Optical AGN: $b$, $r$, $i^{+}$  $z^{++}$, $3.6\mu$m, $4.5\mu$m (784)

\item Normal galaxy population: $v$, $r$, $i^{+}$, $z^{+}$, $3.6\mu$m, $4.5\mu$m
(42,672).
\end{itemize}
We refer the reader to the underlying photometric catalog, \citet{Laigle:2016ku}, for details of the photometry and information for the filters above described above.

\begin{figure}
\centering
\includegraphics[width=0.95\columnwidth]{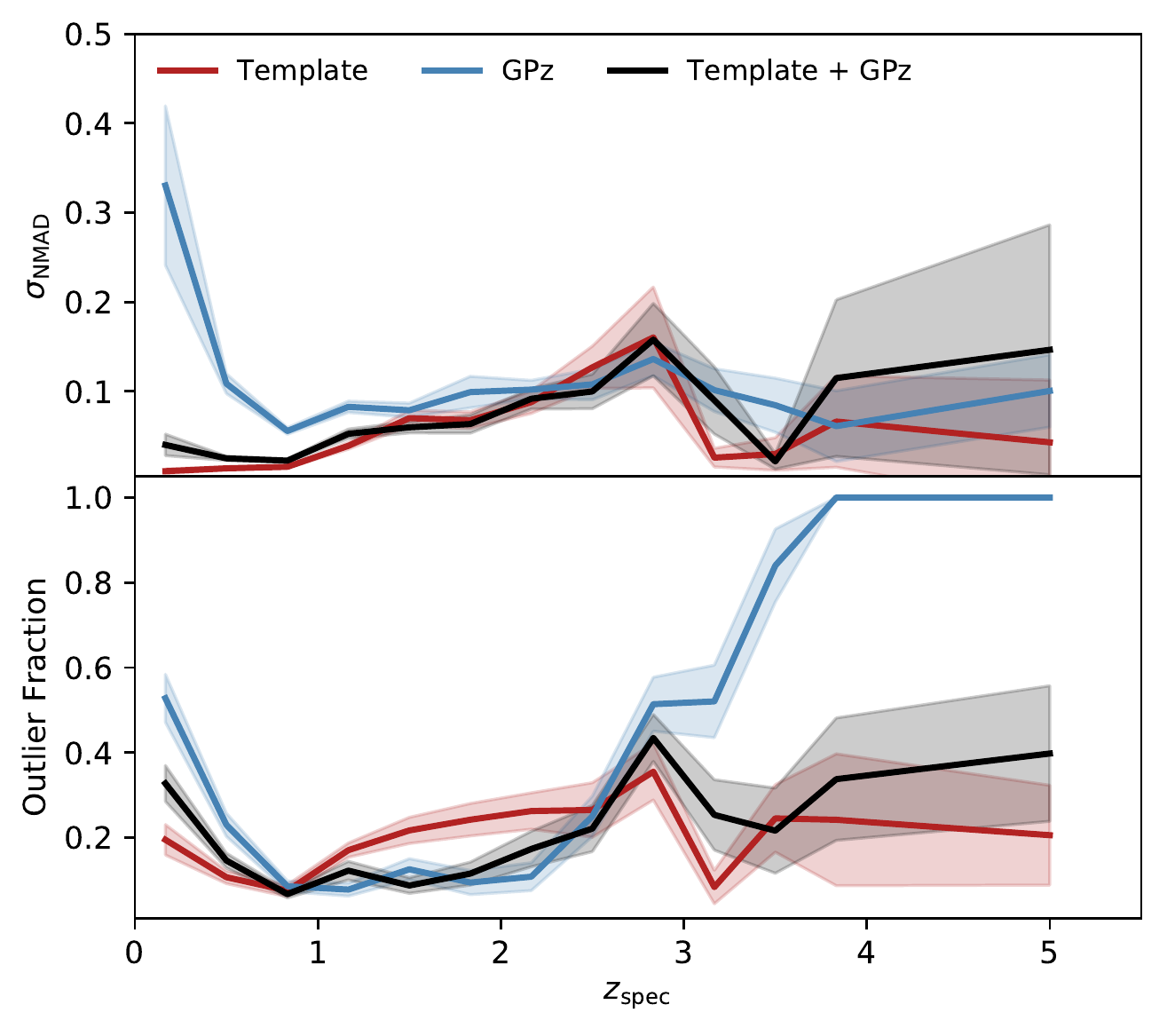}
  \caption{Photometric redshift scatter ($\sigma_{\text{NMAD}}$) and outlier fraction as a function of spectroscopic redshift for AGN sources in the COSMOS field. Lines show the results for sources that pass any of the X-ray/Optical/IR AGN criteria outlined in Section~\ref{sec:data}. Shaded regions around each line represent the standard deviation on the corresponding metric from Bootstrap resampling}
  \label{fig:sigma_vs_z_cosmos}
\end{figure}

Fig.~\ref{fig:sigma_vs_z_cosmos} shows the scatter and outlier fraction for the COSMOS redshifts as a function of redshift for the template only, GPz only and hybrid consensus estimates after Bayesian combination.
Relative to the template estimates, the performance of GPz only consensus is poorer than in Bo\"{o}tes.
Across all redshifts the template estimates have a much lower scatter.
However, between $1 < z < 2.5$ the GPz estimates have a significantly lower outlier fraction than the template only estimates (as for the Bo\"{o}tes, this is the region for which the training sample is least sparse).

Looking at the performance of the COSMOS hybrid estimates, we see a similar behaviour to that observed for the Bo\"{o}tes sample.
In general, the hybrid estimate fairly closely matches the performance of the best individual method - with the scatter comparable to that of the templates estimates across all redshifts but with improved outlier fraction of the GPz estimates at $1 < z < 2.5$.

There are however some regimes where the hybrid estimate does not perform as well as the best individual estimate - notably in the redshift regimes where GPz performs very badly. 
As seen in the previous section, at $z < 1$ and $z > 3$ the GPz estimates become increasingly biased. 
Our earlier conclusion that the bias issues in this regime are primarily due to the sparsity of the training sample are supported by tests on other datasets.
In a forthcoming work, Duncan et al. (in prep), we apply the hybrid photo-$z$ method to over 400 deg$^{2}$ of shallower 'all-sky' data to accompany the release of new LOFAR radio continuum survey data.
Despite the poorer quality optical data, the significantly larger training sample results in much better GPz photo-$z$ estimates at $z < 1$ than either the Bo\"{o}tes or COSMOS samples.
Future applications of the hybrid methodology to deep fields may therefore actually benefit from incorporating additional estimates that use optical data in common with other surveys that may have shallower optical photometry but significantly larger samples upon which to train.

\begin{figure}
 \centering
  \includegraphics[width=0.96\columnwidth]{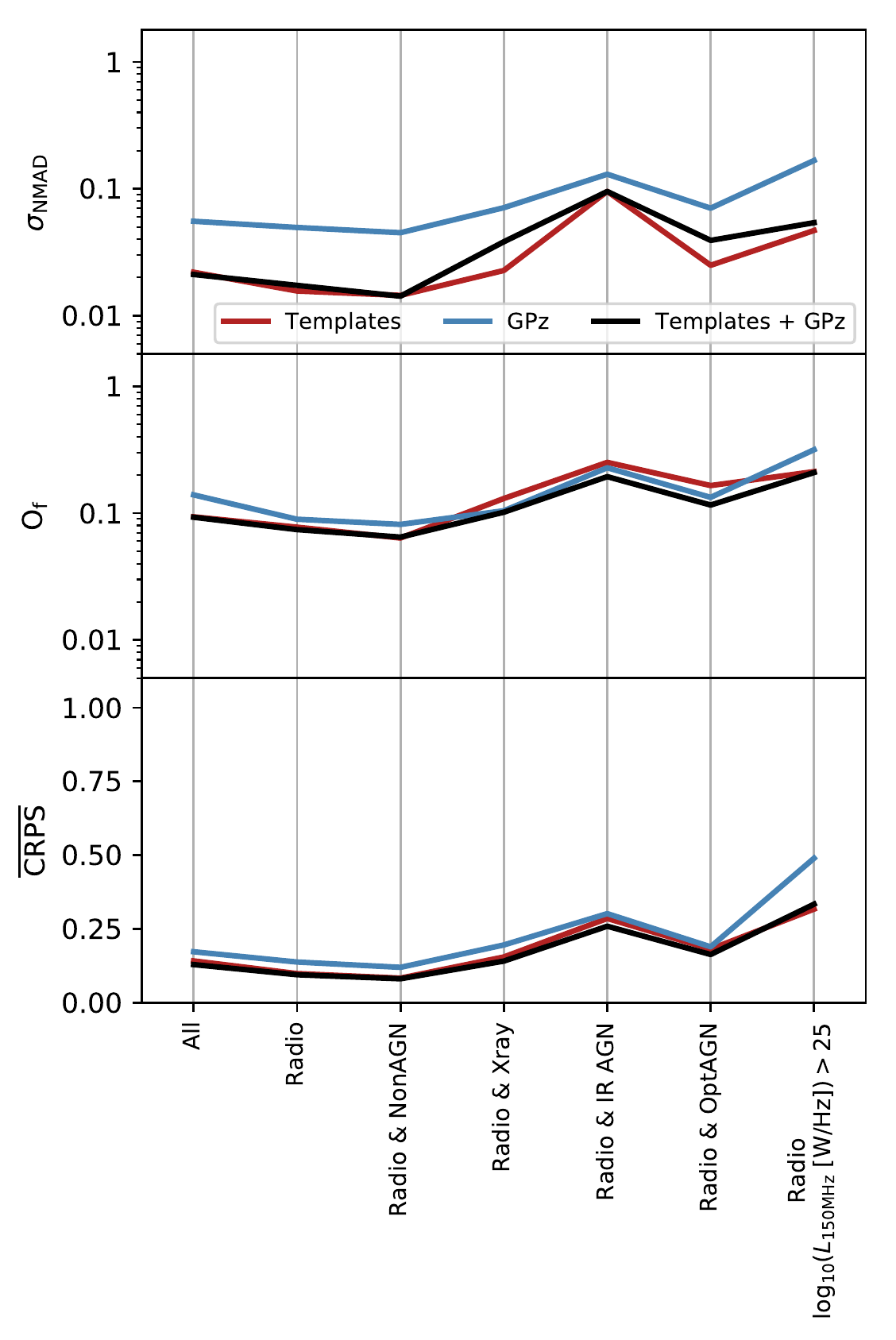}
  \caption{Visualised photometric redshift performance in three metrics ($\sigma_{\textup{NMAD}}$, $O_{\textup{f}}$, $\overline{\textup{CRPS}}$; see Table~\ref{tab:definitions}) for the different COSMOS field radio source subsamples. Compared to the Bo\"{o}tes sample, the performance of the hybrid photo-$z$s compared to the template only estimates is more mixed - with small improvements in scatter and $\overline{\textup{CRPS}}$ for some subsamples but poorer $\sigma_{\textup{NMAD}}$.}
  \label{fig:lin_stats_cosmos}
\end{figure}

Finally, in Fig.~\ref{fig:lin_stats_cosmos} we show the overall performance for different subsamples of the radio population for the new COSMOS estimates.
As expected given the performance statistics as a function of redshift, the hybrid method performs closest to that of the template estimates.
In some subsets of the radio AGN population (X-ray and optical selected AGN), the hybrid method is not able to match or improve upon the scatter.
However it is able to improve upon the outlier fraction and $\overline{\textup{CRPS}}$ by a small margin for these same subsamples.

Overall we conclude that our hybrid methodology can still perform well in deep fields, but the gains to be had over more traditional template fitting are currently much smaller than for typical wide-area surveys.
We note that with the addition of more of the available filters in the field it will be possible to improve the COSMOS GPz estimates.
However, for datasets such as COSMOS it may never be possible for generalised methods such as ours to match the performance of detailed and well curated template estimates for specific subsets of the AGN population \citep[e.g. ][]{Salvato:2008ef,Salvato:2011dq,Marchesi:2016hr}.

\subsection{Prospects and strategies for further improvements}\label{sec:future}
Despite the substantial improvement in photo-$z$ accuracy and reliability for the \textsc{GPz} and hybrid estimates, the inhomogenous photo-$z$ quality across the sub-populations within the radio detected subset indicate that there is still potential for further improvements to be gained.
With regards to the \textsc{GPz} and resulting hybrid estimates, such improvements could potentially come from several different aspects of the methodology.

Firstly, as is the case in all empirical photo-$z$ estimates, the accuracy of \textsc{GPz} is limited by the training sample being used.
Key to the production of accurate photo-$z$s based on training samples is not necessarily the sheer size of the training sample, but rather its ability to fully represent the parameter space probed by the catalogs to which the method will be applied.
The effect of limited training samples can be seen in the performance of \textsc{GPz} at both the very lowest and highest redshifts, the regimes in which the training sample is particularly sparse.
Although our implementation of colour and magnitude based weights within the cost-sensitive learning is able to mitigate some effects of the biased training sample, it will never be able to account for regions of parameter space which are entirely absent from the training data.

In coming years, the problems caused by limited training samples will partly be solved by forthcoming large-scale spectroscopic surveys.
In \citetalias{Duncan:2017wu} we discussed how for the radio-continuum selected population the $>10^{6}$ radio source spectra provided WEAVE-LOFAR \citep{Smith:2016vw} will provide an ideal reference and training sample for photo-$z$ estimates in all-sky radio surveys.
While helpful for improving the template-based estimates, such a training sample will be transformational for machine-learning photo-$z$ estimates of radio sources in future continuum surveys.

In the short term however, it should be possible to better leverage the spectroscopic redshift samples already available in the literature.
The Herschel Extragalactic Legacy Project \citep[HELP:][]{2016ASSP...42...71V} is bringing together all publicly available multi-wavelength datasets within the regions of the sky observed in extragalactic Herschel surveys.
The collation and homogenisation of these many datasets offers the possibility to leverage the extensive spectroscopic datasets in some survey fields to significantly improve estimates in other fields where training samples are particularly sparse.
Furthermore, the inclusion of additional photometric data such as the X-ray flux or radio continuum itself may provide valuable additional information when training empirical estimates.

Secondly, in deeper fields such as Bo\"{o}tes and COSMOS (as opposed to all-sky or large area cosmology surveys), the heterogenous nature of the optical data means that \textsc{GPz} in its current form is not able to make full use of the available information.
This problem is illustrated in Section~\ref{sec:method}, with the only 38.3\% of sources having magnitude information available in five filters and significantly fewer when additional available bands are included.
In the cases where magnitude information is missing as a result of non-detections in the data, training and fitting the photo-$z$s on fluxes rather than magnitudes would largely solve this problem provided the algorithms being used still perform well in the linear regime.
In many other cases however, the missing data can be a result of instrumental effects (e.g. masked regions due to bright stars or diffraction spikes) or differences in the survey coverage.

The flexibility of the hierarchical Bayesian combination procedure outlined in this paper allows for the possibility of training \textsc{GPz} on any/all combinations of the photometric data and combining those estimates to produce a consensus estimate given all the available information.
However, such a procedure would rapidly become impractical in some fields.
Recent developments of the GPz algorithm whereby missing data can be jointly predicted with the redshift (Almosallam et al. in prep) will be of great benefit in the future and could result in significant improvements to the empirical photo-$z$ estimates in these heterogenous deep fields.

Finally, there is also potential for further improvements which can be made to the Bayesian combination.
With additional improvements to the input redshifts themselves, sub-optimal combinations of the various estimates such as those seen at $z\sim3$ in Fig.~\ref{fig:sigma_vs_z} will have less of an effect on the final consensus redshifts.
Nevertheless, more informative priors could be incorporated into the combination procedure which gives more weight to individual estimates in regions of parameter space in which they are known to perform better.
Such an improvement is illustrated in \citet{CarrascoKind:2014jg}, with the performance of Bayesian Model Averaging (BMA) and Bayesian Model Combination (BMC) exceeding that of hierarchical Bayesian combination in their implementation.
However, in the context of photo-$z$s for AGN, we believe these gains will be very small compared to the other strategies outlined in this section.

\section{Summary}\label{sec:summary}

Building on the first paper in this series which explored the performance of template-based estimates \citep[][Paper I]{Duncan:2017wu}, we have presented a study exploring how new estimates from machine learning can be used to significantly improve photo-$z$ estimates for both the radio continuum selected population and the wider AGN population as a whole within the NDWFS Bo\"{o}tes field.
Using the Gaussian process redshift code, GPz, we have produced photo-$z$ estimates targeted at different subsets of the galaxy population - infrared, X-ray and optically selected AGN - as well as the general galaxy population.
The \textsc{GPz} photo-$z$ estimates for the AGN population perform significantly better at $z > 1$ than photo-$z$ estimates produced through template fitting presented in \citetalias{Duncan:2017wu}.
Compared to the template-based photo-$z$s, \textsc{GPz} estimates for the IR/X-ray/Optical AGN population have lower scatter and outlier fractions by up to a factor of four.

By combining these specialised \textsc{GPz} photo-$z$ estimates with the existing template estimates through hierarchical Bayesian combination \citep{Dahlen:2013eu,CarrascoKind:2014jg} we are able to produce a new hybrid consensus estimate that outperforms either of the individual methods across all source types.
The overall quality of photo-$z$ estimates for radio sources that are X-ray sources or optical/IR AGN are vastly improved with respect to \citetalias{Duncan:2017wu}, with outlier fractions and scatter with respect to spectroscopic redshifts reduced by up to a factor of $\sim4$.
When applied to a dataset with deeper photometry and much finer wavelength sampling, we find that the improvement from including \textsc{GPz} is much smaller than for the Bo\"{o}tes sample.
We attribute this effect to the ability of the template estimates to make full use of the increased precision offered by medium or narrow-band photometry.

For both the radio detected population with no strong optical signs of AGN (i.e. radio AGN hosted in quiescent galaxies or star-forming sources) our new methodology also provides significant improvement in the Bo\"{o}tes field.
Despite the template and \textsc{GPz} estimates performing very comparably when treated separately, the combination of the two sets of estimates yields outlier fractions which are a factor of $\approx 2$ lower.
Investigating the new photo-$z$ estimates as a function of radio properties (flux and luminosity), we find that the improvement observed for the radio selected population can likely be attributed to the highest luminosity radio sources for which the \textsc{GPz} estimates (and hence the resulting hybrid estimates) offer huge improvements.

The success of the method despite the small training samples and heterogeneous datasets available is encouraging for future exploitation of deep radio continuum surveys for both the study of galaxy and black hole co-evolution and for cosmological studies.

\section*{Acknowledgements}
The research leading to these results has received funding from the European Union Seventh Framework Programme FP7/2007-2013/ under grant agreement number 607254. This publication reflects only the author's view and the European Union is not responsible for any use that may be made of the information contained therein. KJD and HR acknowledge support from the ERC Advanced Investigator programme NewClusters 321271.




\bibliographystyle{mnras}
\bibliography{library}

\bsp	
\label{lastpage}
\end{document}